\documentclass[preprint,11pt]{JHEP3}

\JHEPspecialurl{http://jhep.sissa.it/JOURNAL/JHEP3.tar.gz}

\usepackage{epsfig,multicol,amsmath}

\usepackage{bm}  
\usepackage{amsmath}     
\usepackage{epsfig,epsf}
\usepackage{rotating}

\usepackage{cite}

\def\beq{\begin{equation}}
\def\eeq{\end{equation}}
\def\bea{\begin{eqnarray}}
\def\eea{\end{eqnarray}}

\def\eq#1{{Eq.~(\ref{#1})}}
\def\fig#1{{Fig.~\ref{#1}}}
\newcommand{\bas}{\bar{\alpha}_S}
\newcommand{\as}{\alpha_S}

\newcommand{\Lb}{\left(}
\newcommand{\Rb}{\right)}
\newcommand{\h}{\frac{1}{2}}
\parskip=\itemsep               

\newcommand{\nn}{\nonumber}
\newcommand{\D}{\partial}

\newcommand{\ga}{\gamma}

\newcommand{\om}{\omega}

\def\pom{{I\!\!P}}

\begin{boldmath}
\title{BFKL Pomeron: modeling confinement}
\end{boldmath}
\author{\Large 
Eugene Levin,${}^{a,b}$ \,and\, Sebastian Tapia${}^{a}$\\
 ${}^a$\, Departamento de F\'\i sica,
Universidad T$\acute{e}$cnica Federico Santa Mar\'\i a   and
Centro Cient\'\i fico-Tecnol$\acute{o}$gico de Valpara\'\i so,
Casilla 110-V,  Valparaiso, Chile\\
${}^b$ \, Department of Particle Physics, School of Physics and Astronomy,
Tel Aviv University, Tel Aviv, 69978, Israel\\
}

\abstract
{In this paper we introduce the confinement into the kernel of the  BFKL equation,
assuming  that the sizes of produced dipoles  cannot be large. The goal of this paper  is to find how this assumption, which leads to a correct exponential decrease of the amplitude at large impact parameters,  affects the main properties of the BFKL Pomeron. We solve the equations for total cross section and $\langle| b^2 |\rangle$ numerically and developed some methods of analytical solutions. The main result is that the modified BFKL Pomeron has the same intercept and $\alpha'_\pom = 0$ as the BFKL Pomeron. It gives us a hope that the unknown confinement will
change only slightly the equations of the CGC/saturation approach.}

\dedicated{PACS: 12.38-t, 12.38.Cy,1 2.38.Lg, 13.60.Hd, 24.85.+p, 25.30.Hm}
\preprint{TAUP  \\
{\tt }\\
\today}

\begin{document}
\section{ Introduction}
The large impact parameter dependence of the scattering amplitude  has been 
the  principle but still unsolved problem in   the  CGC/saturation approach for the past decade. It was shown in Refs.\cite{KW1,KW2,KW3,FIIM} that CGC/saturation approach\cite{GLR,MUQI,MV,REV} that leads to the partial amplitude smaller than unity and satisfies the unitarity constraints, generates the radius of interaction that increases as a power of energy in explicit contradiction to the Froissart theorem\cite{FROI}. It stems from large $b$ behaviour of the BFKL Pomeron\cite{BFKL,LIREV} which has the form: $A\Lb b \gg  1/Q_s\Rb \,\,\propto\,\,s^\Delta/b^2$\footnote{The more detailed discussion of the impact parameter behaviour of the BFKL Pomeron will be done in the next section.}. Amplitude $A\Lb b \gg  1/Q_s\Rb $ becomes of the order of unity at typical $b^2\, \propto\,\,s ^{\Delta}$ leading to $\sigma\,\,\propto\,s^{\Delta}$ in the contradiction to the Froissart theorem ($\sigma\,\,\propto\,\ln^2 s$). The power-like dependence of the scattering amplitude is a direct  consequence of the perturbative QCD technique which is a part of  the CGC/saturation approach. Since the lightest hadron (pion) has a finite mass ($m_\pi$) we know that the amplitude is proportional to $\exp\Lb - 2 m_\pi\,b\Rb$ at large $b$. This exponential behaviour translates into Froissart theorem. Therefore, we have to find how confinement of quarks and gluons being of non-perturbative nature, will change the large $b$ behaviour of the scattering amplitude in the region where this amplitude is small.

This complicated problem in spite of numerous attempts\cite{FIIM,BEST1,BEST2,GBS1,GKLMN,KOLE,HAMU,MUMU}, has not been solved. However we learned several lessons from these  tries. First, in the framework of the DGLAP equation \cite{DGLAP} we can factorize out  the non-perturbative large $b$ behaviour writing for the scattering amplitude\footnote{In this paper we use the following notation: $Y = \ln(1/x)$ where $x$ is the fraction of the energy carried by the dipole, $r$ is the size of scattering dipole, $q$ is the momentum transferred for the scattering amplitude and $\mu_{soft}$ is the scale of soft interaction ($\mu_{soft} \,\sim\,\Lambda_{QCD}$). Notice that $q$ is the Fourier conjugated to the impact parameter $b$.}

\beq \label{SA}
 A\Lb b, Y ,r\Rb \,=\,S\Lb b \Rb A^{DGLAP}\Lb  Y ,r\Rb \eeq
  (see Ref. \cite{GLLMT} ). Indeed, considering the scattering amplitude at fixed  transferred  momentum $q $ ( which is Fourier conjugated to $b$), one can see that for $q \,< \,\mu_{soft}$ the evolutions in $ln(1/r)$ do not depend on $q$. However, for $q\, > \,\mu_{soft}$ the logs take the form $\ln\Lb1/\Lb r q \Rb\Rb$ and the $q$ dependence cannot be absorbed in $S\Lb b \Rb$ in \eq{SA}\cite{GLLMT}.  Using \eq{SA} we can absorbed the non-perturbative corrections at large $b$ in the definition of the saturation scale $Q_s\Lb Y; b\Rb$\cite{MU90,LETU,BRN,ARBR,LGLM,KOLE}.

However, such way of including the non-perturbative large $b$ behaviour does not work\cite{FIIM,BEST1,BEST2,GBS1,GKLMN}  in the case of the BFKL and BK  evolutions  \cite{BFKL,BK}.
 Since we are interested in the behaviour of the scattering amplitude at large $b$ where this amplitude is small, we need to find a way to introduce the non-perturbative corrections directly to the BFKL kernel. Hence the non-linear dynamics does not influence on a solution to this particular problem. We would like to recall that the saturation scale and its dependence on $b$ follows directly from the solution of the BFKL equation (see Ref.\cite{REV} and reference therein).
 It has been checked by numerical calculations (see Refs.\cite{BEST1,BEST2,GBS1,GKLMN}) that if we modify the BFKL kernel introducing {\em by hand} a function that suppressed the production of the dipoles with sizes larger than $1/\mu_{soft}$, the resulting scattering amplitude has the exponential decrease at large impact parameters. 
 
 In this paper we modify the BFKL kernel in the following way:
 \beq \label{MODKER}
 \bas K^{BFKL}\Lb x_{13}, x_{32}| x_{12}\Rb\, \,=\,\,\bas \frac{ x^2_{12}}{x^2_{13}\,x^2_{32}}\,\,\,\,\Longrightarrow\,\,\,\,\as\,
  \frac{ x^2_{12}}{x^2_{13}\,x^2_{32}}e^{ - B \Lb x^2_{13}\,+\,x^2_{32} \Rb}\,\,=\,\, \bas K^{B}\Lb x_{13}, x_{32}| x_{12}\Rb  \eeq
  The motivation for this behaviour of the wave function of a dipole in the confinement region stems from the Gaussian-like form of the wave functions of mesons in holographic AdS/QCD approach\cite{ADSQCD} as well as in the phenomenology of the gluon emission in  at long distances (see Ref.\cite{KOP}). However, we will argue in conclusions that the main results of this paper do not depend on the particular form of \eq{MODKER}.

  Having this modified kernel we are going to answer the following questions :(i) how the intercept of the BFKL Pomeron depends on $B$; (ii)  what is the dependence of $ \langle | b^2 | \rangle $ on Y and the size of the dipole; and (iii) what is dependance of the residue of the BFKL Pomeron on the size of dipole. The goal of this paper to compare the BFKL Pomeron with the modified kernel to  the soft Pomeron we know both from the Regge high energy phenomenology\cite{GLM,KMR} and N=4 SYM with AdS-CFT correspondence
  \cite{AdS-CFT,KOLI,BST1}.  These approaches leads to the soft Pomeron with sufficient large values of the intercept and with the slope ($\alpha'_\pom$)  which is equal to zero (
  $\langle | b^2 | \rangle\,\propto\,\alpha'_\pom \,Y$).

  The result of the paper are the answers to these three questions. We found out that the intercept for the modified BFKL Pomeron is  the same as  the intercept of the BFKL Pomeron with original kernel ($B = 0$ in \eq{MODKER})\footnote{We will 
 call the BFKL Pomeron the solution to the equation with the kernel of \eq{MODKER} with $B=0$.} 
   . At high energies $\langle | b^2 | \rangle \,\rightarrow\, \mbox{Const}$.  In other words we expect that $\alpha'_\pom \to 0$ 
at large $Y$. The Pomeron residue does not depend on the dipole sizes $( r)$ for $r \,<\,1/B$ but it drops for $r \,>\,1/B$.
In short we see that the Pomeron with the modified kernel matches the soft Pomeron as we know it both from N=4 SYM and phenomenology.

The paper is organized as follows.  In the next section we consider the  BFKL Pomeron and discuss its main properties concentrating our attention mostly on the impact parameter dependence. In section 3 we  present the numerical solution for the modified BFKL Pomeron with the kernel of \eq{MODKER} with $B \neq 0$.
In this section we develop several analytical methods to evaluate the intercept of the modified BFKL Pomeron: variational method, semi-classical and diffusion approximations. We solve the equation for $\langle|b^2|\rangle$ and show that the numerical solution and the analytical estimates lead to
$\langle|b^2|\rangle$  which does not depend on energy. In addition, we evaluate the saturation momentum which turns out to show much milder energy behaviour for the modified BFKL Pomeron than for the BFKL equation.

  In conclusions we summarize the results and compare with the soft Pomeron.

\section{Impact parameter dependence of the BFKL Pomeron}


\subsection{The BFKL Pomeron: generalities}

The general solution to the BFKL equation for the scattering amplitude of two dipoles with the sizes $r_1$ and $r_2$ has been derived in Ref.\cite{LIREV} and it takes the form
\bea \label{GENSOL}
&&N\Lb r_1,r_2; Y, b \Rb\,\,= \\
&&\,\,\sum_{n=0}^{\infty}
\int \frac{d \gamma}{2\,\pi\,i}\,\phi^{(n)}_{in}(\gamma; r_2)
\,\,d^2\, R_1 \,\,d^2\,R_2\,\delta(\vec{R}_1 - \vec{R}_2 - 
\vec{b})\,
e^{\omega(\gamma, n )\,Y}
\,E^{\gamma,n}\Lb r_1, R_1\Rb\,E^{1 - \gamma,n}\Lb r_2, R_2 \Rb\nn
\eea
with
\beq \label{OMEGA}
\omega(\gamma, n)\,\,=\,\,\bas \chi(\gamma, n)\,\,  =\,\,\bas \Lb 2 \psi\Lb 1\Rb \,-\,\psi\Lb \gamma + |n|/2\Rb\,\,-\,\,\psi\Lb 1  - \gamma + |n|/2\Rb\Rb;
\eeq
where $~\psi\Lb \gamma\Rb \,\,=\,\,d \ln \Gamma\Lb \gamma\Rb/d \gamma $ and $\Gamma\Lb \gamma\Rb$ is  Euler gamma function. Functions $ E^{n, \gamma} \Lb \rho_{1a},\rho_{2a}\Rb$ are given by the following equations.
\begin{align}\label{EFUN}
  E^{n, \gamma} \Lb \rho_{1a},\rho_{2a}\Rb \,=\, \Lb
  \frac{\rho_{12}}{\rho_{1a} \, \rho_{2a}}\Rb^{1 - \gamma + n/2}
  \, \Lb \frac{\rho^*_{12}}{\rho^*_{1a} \, \rho^*_{2a}}
  \Rb^{1 - \gamma - n/2},
\end{align}
In \eq{EFUN} we use the complex numbers to characterize the point on the plane
\begin{align}
  \rho_i = x_{i,1} + i \, x_{i,2};\,\,\,\,\,\,\, \rho^*_i = x_{i,1} -
  i \, x_{i,2}
\end{align}
where the indices $1$ and $2$ denote  two transverse axes. Notice that
\beq \label{NOT}
\rho_{12}\,\rho^*_{12}\,\,=\,\,r^2_i ;~~~~~~\rho_{1 a}\,\rho^*_{1 a}\,=\,\Lb\vec{R}_i\,-\,\frac{1}{2}\vec{r}_{i}\Rb^2~~~~~~\rho_{2 a}\,\rho^*_{2a}\,=\,\Lb\vec{R}_i\,+\,\frac{1}{2}\vec{r}_{i}\Rb^2
\eeq
At large values of $Y$ the main contribution stems from the first term with $n =0$.  For this term \eq{EFUN} can be re-written in the form
\beq \label{E}
E^{\gamma,0}\Lb r_i,R_i\Rb \,\,=\,\,\left( \,\frac{r^2_{i}}{(\vec{R}_i
\,+\,\frac{1}{2}\vec{r}_{i})^2\,\,
(\vec{R}_i\,-\,\frac{1}{2}\vec{r}_{i})^2}\,\right)^{1 - \gamma}\,\,.
\eeq

The integrals over $R_1$ and $R_2$ were taken in Refs.\cite{LIREV,NAPE} and at $n=0$ we have
\bea \label{H}
&&H^\ga\Lb w, w^*\Rb\,\,\equiv\,\,\int d^2\,R_1\,E^{\ga,0}\Lb r_{1},R_1\Rb,\, E^{1 - \ga, 0}\Lb r_{2}, \vec{R}_1
\,-\,\vec{b}\Rb\,= \\
&&
\,\frac{ (\gamma - \h)^2}{( 
\gamma (1 - \gamma)
)^2} \Big\{b_\ga\,w^\gamma\,{w^*}^\gamma\,F\Lb\gamma, \gamma, 2\gamma, w\Rb\,
F\Lb\gamma, \gamma, 2\gamma, w^*\Rb
\,+ \nn\\
&&  b_{1 - \ga} w^{1 -
\gamma}{w^*}^{1-\gamma}
F\Lb 1 - \gamma, 1 -\gamma, 2 - 2\gamma, w\Rb\,F\Lb 1 - \gamma,1 -\gamma,2 
-2\gamma, w^*\Rb \Big\}\nn
\eea
where $F$ is hypergeometric function \cite{RY}. In  \eq{H}
$w\,w^*$ is equal to
\beq \label{W}
w\,w^*\,\,=\,\,\frac{r^2_{1}\,r^2_{2,t}}{\Lb\vec{b} - \h\Lb\,\vec{r}_{1}\,
- \,\vec{r}_{2}\Rb\Rb^2
\,\Lb\vec{b} \,+\, \h \Lb\,\vec{r}_{1} \,- \,\vec{r}_{2}\Rb\Rb^2}
\eeq
and  $b_\ga$ is equal to
\beq \label{BGA}
b_{\ga} \, = \, \pi^3 \, 2^{4(1/2 - \ga)} \, \frac{\Gamma \Lb\ga \Rb}{\Gamma \Lb 1/2 - \ga \Rb}
  \, \frac{\Gamma \Lb 1 - \ga  \Rb}{\Gamma \Lb 1/2 + \ga \Rb}.
\eeq

Finally, the solution at large $Y$ takes the form
\beq \label{FINSOL}
N\Lb r_1,r_2; Y, b \Rb\,\,= \,\,
\int \frac{d \gamma}{2\,\pi\,i}\,\phi^{(0)}_{in}(\gamma; r_2)
\,
e^{\omega(\gamma, 0 )\,Y}\,H^\ga\Lb w, w^*\Rb
\eeq

\eq{FINSOL}  shows that at large $b\, \gg\, r_1 $ and $r_2$ $w
\,w^*\, = \,r^2_1r^2_2/b^4\,\ll\,1$. Therefore,  we can replace $F$ functions in \eq{H} by unity and \eq{FINSOL} degenerates to the following expression
\beq \label{LABEXP}
N\Lb r_1,r_2; Y, b \Rb\,\,= \,\,
\int \frac{d \gamma}{2\,\pi\,i}\,\phi^{(0)}_{in}(\gamma; r_2)
\,
e^{\omega(\gamma, 0 )\,Y}\Big\{b_\ga \Lb \frac{r^2_1\,r^2_2}{b^4}\Rb^{\ga} 
\,+  b_{1 - \ga}  \Lb \frac{r^2_1\,r^2_2}{b^4}\Rb^{1 - {\ga}}  \Big\}\,\,\,\longrightarrow\,\,\,\,\frac{r_1\,r_2}{b^2}\, e^{\omega_0 \,Y} 
\eeq
where at $Y \gg 1$  $\ga \to 1/2$ and $\omega_0 \,=\,\bas \chi(1/2)$. One can see that $ N\Lb r_1,r_2; Y, b \Rb\,<\,1$ for $b^2 \,\leq \,r_1 r_2 e^{\omega_0 Y}$\cite{KW1,KW2,KW3}.

For DGLAP evolution the essential $r_1 ~\ll~r_2$ and $\gamma \to 0 $ and \eq{FINSOL} takes the form
\beq \label{DLA}
N\Lb r_1,r_2; Y, b \Rb\,\,= \,\,
\int \frac{d \gamma}{2\,\pi\,i}\,\phi^{(0)}_{in}(\gamma; r_2)
\,
e^{\omega(\gamma, 0 )\,Y} \Lb w\,w^*\Rb^\ga
\eeq
One can see that or $b \ll r_2$ $w\,w^*\,=\,r^2_1/r^2_1$ and the impact parameter dependence can be introduced through non-perturbative initial condition. However, for $ r_1 \,\ll\, |\vec{b} - \h \vec{r}_2| \,\,\ll\,\,r_2$ $w\,w^*\,=\,r^2_1/ |\vec{b} - \h \vec{r}_2|^2$ and $b$ dependence cannot be absorbed in the initial condition.

\begin{boldmath}
\subsection{Equation for $\langle|b^2\Lb Y,l\Rb|\rangle$ }
\end{boldmath}

In  this section we derive the equation for  $\langle|b^2\Lb Y,l\Rb|\rangle$  defined as
\beq \label{B2}
 \langle|b^2\Lb Y,l\Rb|\rangle\,\,\,=\,\,\,\frac{\int d^2 b\, b^2\, N^{BFKL}\Lb r_1,r_2; Y, b \Rb}{\int d^2 b\, N^{BFKL}\Lb r_1,r_2; Y, b \Rb}
\eeq

The BFKL equation takes the form:
\bea \label{BFKL}
\hspace{-0.8cm}\frac{\partial N^{BFKL}\Lb x_{12}, b; Y \Rb}{\partial Y} &=&
\bas \int d^2 x_{13} \,\frac{x^2_{12}}{x^2_{13}\,x^2_{32}}\,\Big\{ 2\,N^{BFKL}\Lb x_{13},\vec{b} - \h \vec{x}_{32}; Y \Rb  \,\,-\,\, N^{BFKL}\Lb x_{12}, b; Y \Rb\Big\}\nn\\
&=&\bas \int d^2 x_{13} \,\frac{1}{x^2_{32}}\,\Big\{ 2\,\widetilde{N}^{BFKL}\Lb x_{13},\vec{b} - \h \vec{x}_{32}; Y \Rb  \,\,-\,\,\frac{x^2_{12}}{x^2_{13}} \widetilde{N}^{BFKL}\Lb x_{12}, b; Y \Rb\Big\}
\eea
where $\widetilde{N}^{BFKL}\Lb x_{12}, b; Y \Rb \,\,=\,\,N^{BFKL}\Lb x_{12}, b; Y \Rb/x^2_{12}$ and $x_{12}$ is the size of the dipole ($r_1$ in the notation of the previous section). The size of the second scattered dipole $r_2$ we suppress in the notation.

 Integrating \eq{BFKL} over the impact parameter we obtain the equation for
${\cal N}^{BFKL}\Lb x_{12}; Y \Rb~=$\\$\int d^2 b\,\widetilde{N}^{BFKL}\Lb x_{12},b ; Y \Rb$ which takes the form
\beq \label{EQ1}
\frac{\partial {\cal N}^{BFKL}\Lb x_{12}; Y \Rb}{\partial Y}\,\,=\,\,\,\bas \int d^2 x_{13} \, \frac{1}{x^2_{32}}\,\Big\{ 2\,{\cal N}^{BFKL}\Lb x_{13}; Y \Rb  \,\,-\,\,\frac{x^2_{12}}{x^2_{13}}{\cal N}^{BFKL}\Lb x_{12}; Y \Rb\Big\}
\eeq

Multiplying \eq{BFKL} by $b^2$ we derive the equation for ${\cal \widehat{N}}^{BFKL}\Lb x_{12}; Y \Rb = \int d^2 b\,b^2\,\widetilde{N}^{BFKL}\Lb x_{12},b ; Y \Rb$:
\bea 
&&\frac{\partial {\cal \widehat{ N}}^{BFKL}\Lb x_{12}; Y \Rb}{\partial Y}\,\,= \label{EQ2}\\
&&=\,\,\bas \int \!\!\!\!\int d^2 b' \, d^2 x_{13}\,\frac{1}{x^2_{13}}\Big(\vec{b'} + \h \vec{x}_{32}\Big)^2\frac{1}{x^2_{32}}\,\Big\{ 2\,\widetilde{N}^{BFKL}\Lb x_{13},\vec{b'} \equiv\vec{b} - \h \vec{x}_{32}; Y \Rb  \,\,-\,\,\frac{x_{12}}{x^2_{13}} \widetilde{N}^{BFKL}\Lb x_{12}, b; Y \Rb\Big\}\nn\\
&&=\,\,\,\bas \int   d^2 x_{13} \,\frac{1}{x^2_{32}}\,\Big\{ 2\,{\cal\widehat{N}}^{BFKL}\Lb x_{13}; Y \Rb  \,\,-\,\,\frac{x_{12}}{x^2_{13}} {\cal \widehat{N}}^{BFKL}\Lb x_{12}; Y \Rb\Big\} \,\,+\,\,\h
 \bas \int d^2 x_{13} \, {\cal N}^{BFKL}\Lb x_{13}; Y \Rb\nn\\
 &&+\,\,\Bigg\{\h \bas \int  \!\!\!\!\int d^2 b' \, d^2 x_{13} \,\,\vec{b'}\cdot\vec{x}_{32}\,\frac{1}{x^2_{32}}\, 2\,\widetilde{N}^{BFKL}\Lb x_{13},\vec{b'} \equiv\vec{b} - \h \vec{x}_{32}; Y \Rb\,\,\,=\,\,\,0\Bigg\}\label{BX}
 \eea
The last term (see \eq{BX} ) is equal to 0. It  follows directly from \eq{FINSOL} and the expression for $w w^*$ of \eq{W},
since they show that $\widetilde{N}^{BFKL}\Lb x_{13},\vec{b'}; Y \Rb$ is even function of $\vec{b'}$ ($ 
\widetilde{N}^{BFKL}\Lb x_{13},\vec{b'}; Y \Rb\,=\,\widetilde{N}^{BFKL}\Lb x_{13}, - \vec{b'}; Y \Rb$). Therefore, the integral over $b'$ in \eq{BX} vanishes.

We need to solve \eq{EQ1} and \eq{EQ2} to find  $ \langle|b^2\Lb Y,l\Rb|\rangle\,\, =\,\,{\cal \widehat{ N}}^{BFKL}\Lb x_{12}; Y \Rb\Big{/}{\cal N}^{BFKL}\Lb x_{12}; Y \Rb$.  The initial conditions for these equations are taken in the form
\beq \label{IC}
{\cal N}^{BFKL}\Lb x_{12}; Y =0\Rb\,\,=\,\,\,\ln\Lb1/\Lb x^2_{12}\Lambda^2_{QCD}\Rb\Rb;~~~~~~~~~~
{\cal \widehat{ N}}^{BFKL}\Lb x_{12}; Y =0\Rb \,\,=\,\,b^2_0\,\ln\Lb 1/\Lb x^2_{12}\Lambda^2_{QCD}\Rb\Rb;
\eeq
where $b^2_0$ is the value of $ \langle| b^2 |\rangle\,$ at $Y=0$.

\eq{EQ1} has a well known  solution:
\beq \label{SOL1}
{\cal N}^{BFKL}\Lb x_{12}; Y \Rb\,\,=\,\,\int^{\epsilon + i \infty}_{\epsilon - i \infty} \frac{d \nu}{2 \pi}\,\frac{1}{\Lb i \nu - \h \Rb^2}e^{\omega\Lb \nu \Rb\,Y\,\,+\,\,\Lb - \h + i \nu\Rb \,l}
\eeq
where $l\,=\,\ln\Lb x^2_{12}\,\Lambda^2_{QCD}\Rb$ and  $\omega\Lb \nu\Rb\,=\,\omega\Lb\h + i \nu, 0\Rb$ (see \eq{OMEGA}).

At large $Y$ the main contribution stems from $\nu \to 0 $ where we can use the simplified form of  BFKL kernel $\om\Lb \ga,0\Rb$:  its expansion at small values of $\nu$ (diffusion approximation)
 \beq 
 \label{DIFF}
 \om\Lb \nu \Rb\,\,=\,\,\om_0\,\,-\,\,D_0 \,\nu^2 \,\,\,\,\,\,\mbox{with}\,\,\,\,\,\,\om_0 \,=\,4 \ln2 \, \bas  = 2.772 \, \bas \,\,\,\mbox{and}\,\,\,\,D_0\,=\,14 \zeta\Lb 3 \Rb\,  \bas \,=\,16.828 \, \bas
 \eeq

Taking integral over $\nu$ we have

\beq \label{SOLDIFF}
{\cal N}^{BFKL}_{diff}\Lb x_{12}; Y \Rb\,\,=\,\,\frac{4}{\sqrt{4 \,D_0\,Y}} \,e^{\omega_0\,Y\,  \, - \,\,\h \,l \,-\,\frac{l^2}{4 D_0 Y}}
\eeq

The second term in \eq{EQ2} is a function of only $Y$.  We solve \eq{EQ2} using double Mellin transform
 
 \beq \label{MELLIN}
 {\cal \widehat{ N}}^{BFKL}\Lb x_{12}; Y \Rb\,\,=\,\, \int^{\epsilon + i \infty}_{\epsilon - i \infty}\frac{d \omega}{ 2 \pi i} \int^{i\epsilon +  \infty}_{i\epsilon - \infty} \widehat{n}\Lb \om, \nu\Rb\,e^{\omega \,Y \,+\,\Lb - \h + i \nu\Rb l}
 \eeq
 Plugging  \eq{MELLIN} in \eq{EQ2} we obtain that
 \bea
 \label{EQ2ON}
 \Lb \om \,-\,\om\Lb\nu\Rb\Rb \widehat{n}\Lb \om, \nu\Rb\,\,&=&\,\,\h \bas \frac{\pi\,b^2_0}{\Lb\h - i \nu\Rb} \int \frac{d \nu'}{2 \pi i}
 \frac{1}{\om \,-\,\om\Lb \nu'\Rb} \,\frac{1}{\Lb  \h  - i \nu'  \Rb^2}\,\int dl\,e^{ \Lb  \h + i \nu'\Rb l}\nn\\
  &=&   \,\,\h \bas \frac{\pi\,b^2_0}{\Lb \h - i \nu\Rb} \int \frac{d \nu'}{2 \pi i}
 \frac{1}{\om \,-\,\om\Lb \nu'\Rb} \,\frac{1}{\Lb i \nu' - \h \Rb^2}\,\frac{1}{  \h + i \nu'}
 \eea
 The l.h.s. of \eq{EQ2ON} is the result of the integration over $x_{13}$ in \eq{EQ2}.  This integration demonstrates that in the framework of the BFKL equation we cannot find ${\cal \widehat{ N}}^{BFKL}\Lb x_{12}; Y \Rb$ since the integral is divergent. It is expected since the BFKL equation is conformal symmetric and therefore, the dimensional scale can be originated only in  the initial condition. For example, in dipole-dipole scattering the natural scale is the size of the target dipole ($b_0$). In this case, the choice of $l$ is $l\,=\,\ln\Lb x^2_{12}/b^2_0\Rb$ which is used in \eq{EQ2ON} and the initial condition of \eq{IC} can be written as 
 \beq \label{ICBFKL}
 {\cal \widehat{ N}}^{BFKL}\Lb x_{12}; Y =0\Rb \,\,=\,\,b^2_0\,\ln\Lb b^2_0/\Lb x^2_{12}\Rb\Rb;
 \eeq
  
  As one can see in \eq{EQ2ON} we integrated over $l \,<\,0$ and, thereby, introduced the infrared cutoff,  assuming that the dipole size is less than $b_0$.  Hence  \eq{EQ2ON} includes a modification of the BFKL equation, which we apply only for the dipoles with sizes smaller than $b_0$. For the modified BFKL equation such a  cutoff  is an intrinsic property of the equation. Therefore, introducing the cutoff in \eq{EQ2ON} we model the modified BFKL equation with the kernel of \eq{MODKER} with the goal to obtain the analytical solution which we cannot hope to get for the modified kernel of
 \eq{MODKER}. It should be noticed that the factor  $1/{\Lb \h - i \nu\Rb} $ stems from the condition that the  non homogeneous term in \eq{EQ2ON} we consider only for $l < 0$.

 Generally speaking we can add to solution of \eq{EQ2ON} any solution of the homogeneous equation in the form
 
 \beq \label{ADD}
 \Delta  \widehat{n}\Lb \om, \nu\Rb\,\,=\,\,  n\Lb \nu\Rb\frac{1}{\om \,-\,\om\Lb \nu \Rb}
 \eeq
 with arbitrary $n\Lb \nu\Rb$.
 
 $\om\Lb \nu\Rb$ has pole of the first order at $i \nu \to \pm 1/2$ and at large values of $Y$ the main contribution stems from $\nu \to 0$. Using this features of $\om\Lb \nu\Rb$ and its expansion at small values of $\nu$ (see \eq{DIFF})
 we can take the integral over $\nu'$ in \eq{EQ2ON}. It takes the form
\beq \label{SOL21}
\h \pi \bas \frac{b^2_0}{\Lb \h - i \nu\Rb} \,\Lb\frac{1}{\bas}\,\,+\,\,\frac{4}{\sqrt{D_0 \Lb \om\,-\,\om_0 \Rb}}\Rb
\eeq
Using \eq{SOL21} the general solution can be written as
\beq \label{SOL22}
\widehat{n}\Lb \om, \nu\Rb\,\,\,=\,\,\,\frac{1}{\om\,-\,\om\Lb \nu \Rb}\,\frac{b^2_0}{\Lb \h - i \nu\Rb}\,\Bigg\{\Lb \frac{\pi}{2}\,\,+\,\,\frac{2\pi \bas}{\sqrt{D_0 \Lb \om\,-\,\om_0\Rb}}\Rb\,\,-\,\,n\Lb \nu \Rb \Bigg\}
\eeq
 Function $n\Lb \nu\Rb$ should be found from the initial condition
 \beq \label{SOL23}
  \int^{\epsilon + i \infty}_{\epsilon - i \infty}\frac{d \omega}{ 2 \pi i}
  \,\, \frac{1}{\om\,-\,\om\Lb \nu \Rb}\,\,\Bigg\{\Bigg( \frac{\pi}{2}\,\,+\,\,\frac{2\pi \bas}{\sqrt{D_0 \Lb \om\,-\,\om_0 \Rb}}\Bigg)\,\,-\,\,n\Lb \nu \Rb \Bigg\}\,\,=\,\,\frac{1}{\h -  i \nu}
  \eeq
 Resolving \eq{SOL23} we obtain
 \beq \label{SOL24}
 n\Lb \nu \Rb\,\,=\,\,\frac{\pi}{2} - \frac{1}{\h - i \nu} \,- \frac{2 \pi \bas}{ D_0 \nu}
 \eeq
Finally
\beq \label{SOL25}
 \widehat{n}\Lb \om, \nu\Rb\,\,\,=\,\,\,\frac{1}{\om\,-\,\om\Lb \nu \Rb}\,\frac{b^2_0}{\Lb \h - i \nu\Rb}\,\Bigg\{\frac{1}{\h - i\nu} \,\,+\,\,\frac{2\pi \bas}{\sqrt{D_0 \Lb \om\,-\,\om_0 \Rb}}\,\,+\,\,\frac{2 \pi \bas}{ D_0 \nu}\Bigg\}
 \eeq
 
 Taking integral over $\om$ we get for $\nu \to 0$
 \beq \label{SOL26}
   \int^{\epsilon + i \infty}_{\epsilon - i \infty}\frac{d \omega}{ 2 \pi i} \widehat{n}\Lb \om, \nu\Rb\,\,=\,\,   
   \frac{b^2_0}{\Lb \h - i \nu\Rb}e^{ \om\Lb \nu \Rb\,Y}  \left\{\Lb \frac{1}{\h - i \nu} +\frac{2 \pi \bas}{D_0\,\nu}\Rb  - \frac{2 \bas \pi }{D_0 \nu}\,\mbox{Erfc}\Lb\sqrt{D_0 \nu^2 Y}\Rb \right\}
   \eeq
   One can see that at small $D_0 \nu^2 Y$ the factor in curly bracket vanishes. Taking the integral over $\nu$ using the steepest decent method we see that this parameter is equal to $D_0 \nu_{SP}^2 Y  \, =\, l^2/\Lb 4 D_0 Y\Rb$. Since we are interested mostly in the solution at $l $ which is not very large concentrating our effort on so called Regge domain, we can safely consider this parameter as being small.  Expanding $\mbox{Erfc}\Lb\sqrt{D_0 \nu^2 Y}\Rb $ with respect to this parameters, we see that
   $\Big\{ \dots\Big\}\,\,=\,\,  1\,\,\,+\,\,\Lb 2 \bas \sqrt{\pi}/\sqrt{D_0}\Rb \sqrt{Y}$.
   
   Therefore, plugging solution of \eq{SOL26} into \eq{B2} one can see that
   \beq \label{SOL27}
   \langle|b^2|\rangle \,=\,\,b^2_0\, \Big( 1\,\,\,+\,\,\frac{2 \bas \sqrt{\pi}}{\sqrt{D_0}}\,\sqrt{Y}\Big)
   \eeq
   in this approximation.  Such a behaviour of $  \langle|b^2|\rangle$ versus $Y$ has been expected (see Ref.\cite{B2EXPE}). Comparing this behaviour with
   $\langle |b^2|\rangle = 4 \alpha'_\pom\,Y$ one can see that $\alpha'_\pom$ for the considered modification of the BFKL Pomeron is equal to zero.

     \begin{figure}[ht]
    \begin{tabular}{c c}
      \includegraphics[width=8.5cm]{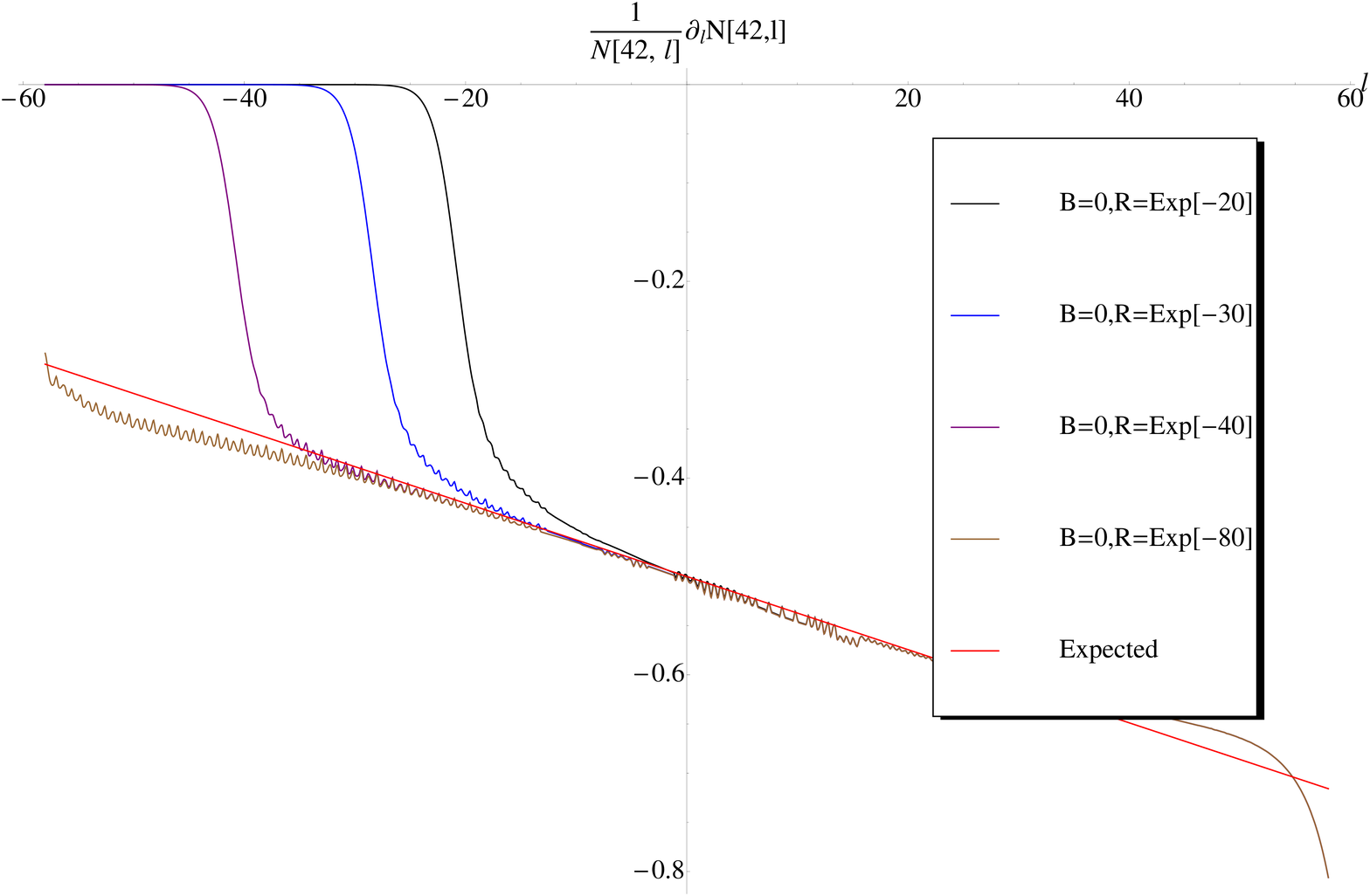} &  \includegraphics[width=8.5cm]{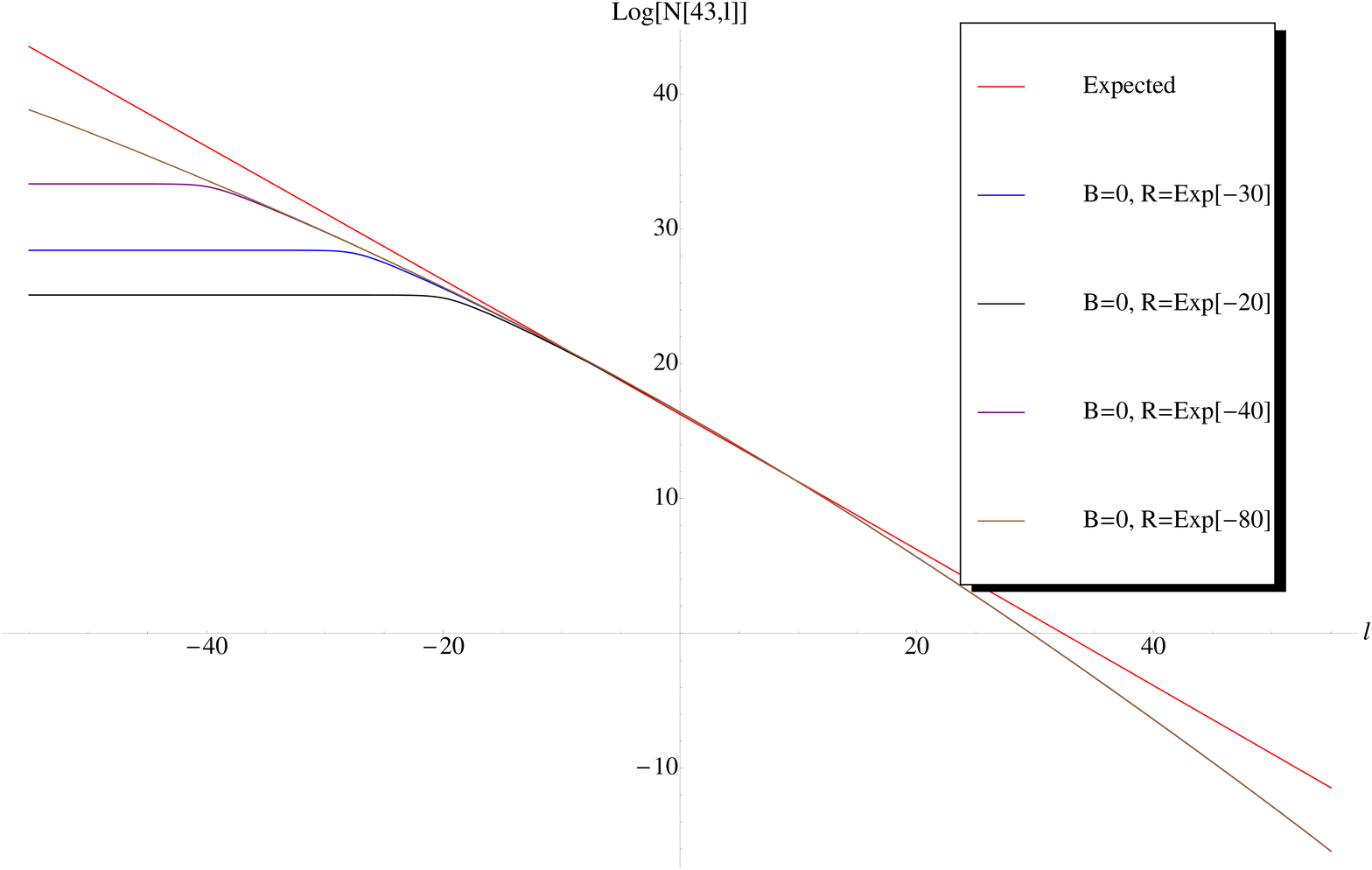} \\
      \fig{b0}-a & \fig{b0}-b\\
      \end{tabular}
      \caption{ $\partial \ln \Big({\cal N}^{BFKL}\Lb l; Y \Rb\Big)/ \partial l$ (see \protect\fig{b0}-a) and $  \ln \Big({\cal N}^{BFKL}\Lb l; Y \Rb\Big)$ (see \protect\fig{b0}-b)         for the BFKL equations (see \protect\eq{EQ1} and \protect\eq{EQ2} with the BFKL kernels) as function of $l$ at $Y = 43$. $\bas$ is chosen to be equal to 0.2  and  $l = \ln\Lb x^2_{12}\,\Lambda^2_{QCD}\Rb $. Curves correspond to different values of $R$. The red curve shows the solution of \protect\eq{SOLDIFF}.
      }
\label{b0}
   \end{figure}


\subsection{Numerical solution}
Searching for numerical solution of \eq{EQ1} we introduce the regulator at short distances $R$ in the following way
\beq \label{SOLNUM1}
\frac{\partial {\cal N}^{BFKL}\Lb x_{12}; Y \Rb}{\partial Y}\,\,=\,\,\,\bas \int d^2 x_{13} \, \frac{1}{x^2_{32} + R^2}\,\Big\{ 2\,{\cal N}^{BFKL}\Lb x_{13}; Y \Rb  \,\,-\,\,2\,\frac{x^2_{12}}{x^2_{13}\,+\,x^2_{23}\,+\,2\,R^2}{\cal N}^{BFKL}\Lb x_{12}; Y \Rb\Big\}
\eeq
We solve \eq{SOLNUM1} at fixed but different $R$ making it smaller until the answer will not depend on $R$.
We use the solution given by \eq{SOLDIFF} ,   as a check of the accuracy of our numerical calculations.  The procedure of the derivation of \eq{SOLNUM1} from \eq{EQ1} is standard and it is described in Ref.\cite{REV} for example.

\fig{b0} shows the numerical solutions of the BFKL equations (see \eq{EQ1} and \eq{SOLNUM1}) for $\partial \ln \Big({\cal N}^{BFKL}\Lb l; Y \Rb\Big)/ \partial l$ (see \protect\fig{b0}-a) and $  \ln \Big({\cal N}^{BFKL}\Lb l; Y \Rb\Big)$ (see \protect\fig{b0}-b)    as functions of $l  = \ln\Lb x^2_{12}\,\Lambda^2_{QCD}\Rb$
for different values of the short distance regulator $R$ at $Y = 43$. Comparing these curves with the prediction of the solution of \eq{SOLDIFF} shown in \fig{b0} by the red curve, one can see that  our numerical problems are concentrated in the region of very short distances where the value of $x_{12} $ approaches $R$. However, we see that in the region $ -20 < l < 20$ the numerical solution coincides with \eq{SOLDIFF} with good accuracy.
     
     \begin{figure}[ht]
    \begin{tabular}{c c}
      \includegraphics[width=8.5cm]{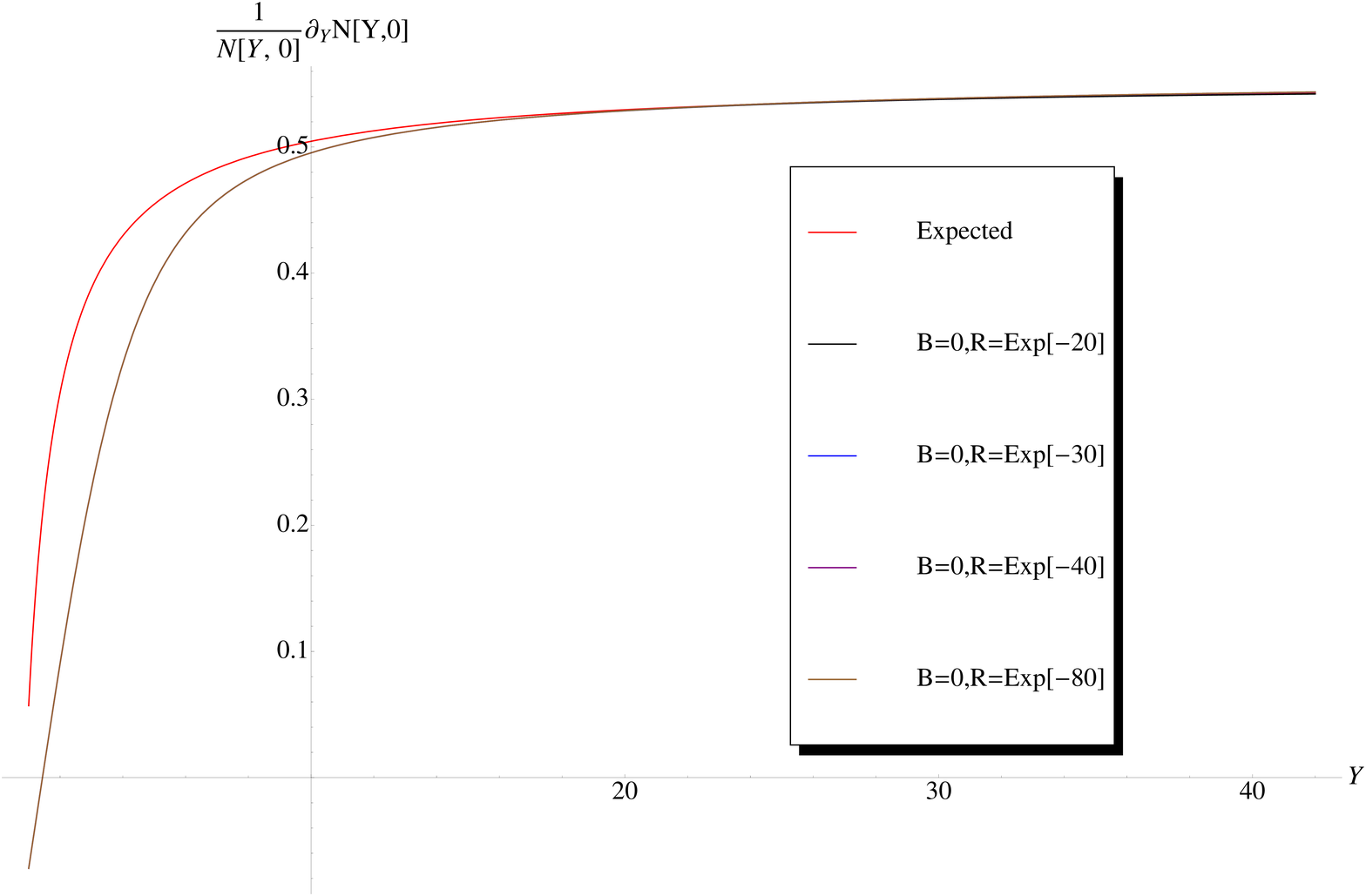} &  \includegraphics[width=8.5cm]{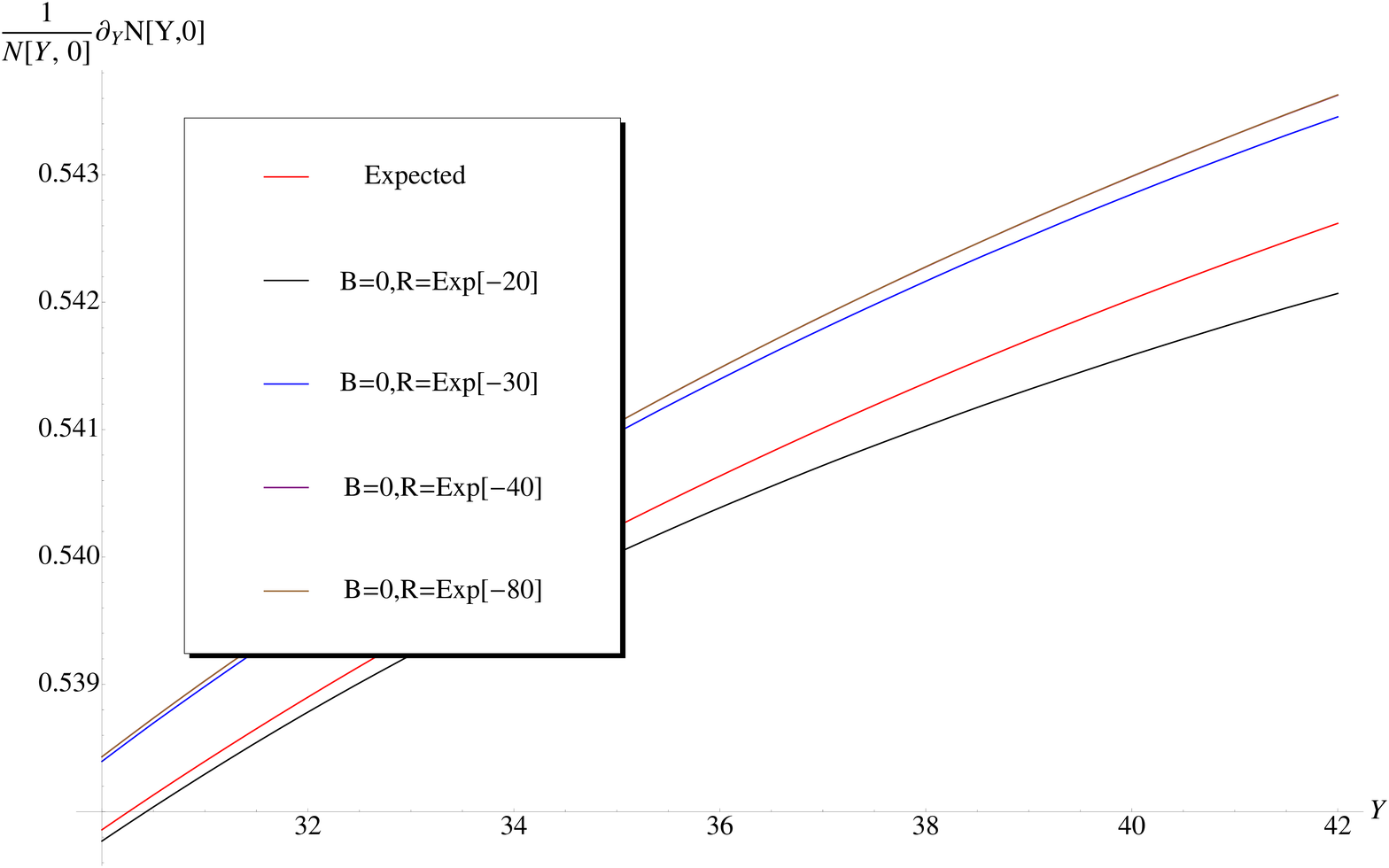} \\
      \fig{DY}-a & \fig{DY}-b\\
      \end{tabular}
      \caption{ $\partial \ln \Big({\cal N}^{BFKL}\Lb l; Y \Rb\Big)/ \partial Y$ (see \protect\fig{DY}-a  and the zoomed picture in  \protect\fig{DY}-b)         for the BFKL equations (see \protect\eq{EQ1} \protect\eq{SOLNUM1} with the BFKL kernels) as function of $Y$ at $l = 0$. $\bas$ is chosen to be equal to 0.2  and  $l = \ln\Lb x^2_{12}\,\Lambda^2_{QCD}\Rb $. Curves correspond to different values of $R$. The red curve shows the solution of \protect\eq{SOLDIFF}.
      }
\label{DY}
   \end{figure}

 In \fig{DY} we plotted $\partial \ln \Big({\cal N}^{BFKL}\Lb l; Y \Rb\Big)/ \partial Y$ as a function of $Y$ for different values of $R$. Comparing the curves for different $R$ with the solution of \eq{SOLDIFF} shown by red line in \fig{DY} one can conclude: first, that the value of the BFKL  Pomeron intercept does not depend on the value of $R$ and, second, that the numerical solution reproduces  quite well the analytical one, given by \eq{SOLDIFF} . However, \fig{DY}-b shows that our numerical value for the Pomeron intercept systematically above the analytical one approximately on 
 0.001 which we consider as a systematic error of our calculations.
  
\section{Modified BFKL Pomeron}
 In this section we solve \eq{EQ1} and \eq{EQ2} in which the BFKL kernel are replaced by $K^{B}\Lb x_{13},x_{32}|x_{12}\Rb$. As has been mentioned above the numerical calculations in Refs.\cite{BEST2,BEST1,GBS1,GKLMN} show that  such a modification of the BFKL kernel leads the exponential decrease of the scattering amplitude at large values of the impact parameter.  Actually, we can see this directly from the equation (see \eq{BFKL}) . Indeed, one can see that the main contribution at large $b$ stems from the region where $|\vec{b} - \vec{x}_{32}| \,\leq\,x_{12}$. At such $x_{32}$ the equation takes the form
 \beq \label{MBFKL1}
 \frac{\partial N\Lb x_{12} \approx 2 b , b; Y \Rb}{\partial Y} \,\,=\,\,
4 \bas \int d^2 x_{13} \,\frac{1}{4 b^2}e^{ - 4\,B\, b^2 - B\,x^2_{13}} \, 2\,N\Lb x_{13},0; Y \Rb  ~~ \propto ~~e^{ - 4\,B\,b^2}
\eeq
  
We can replace $ \exp\Lb - B\,x^2_{13}\Rb$ by $K_0\Lb \mu x_{13}\Rb$ to reproduce correct $\exp\Lb - \mu \,b\Rb$
at large $ b$ (see more detailed analysis of the form of the BFKL kernel in Ref.\cite{BEST1}). Nevertheless, we solve the equation with the kernel of \eq{MODKER} since we interested in the behaviour of the   $\langle|b^2\Lb Y,l\Rb|\rangle$  versus energy for which the particular form of the kernel is not important. 

The initial condition for solving  \eq{EQ1} and \eq{EQ2} we take  in the following form
\beq \label{IC1}
{\cal N}^{BFKL}\Lb x_{12}; Y =0\Rb\,\,=\,\,\ln\Lb1/\Lb x^2_{12}\,B\Rb\Rb;~~~~~~~~~~
{\cal \widehat{ N}}^{BFKL}\Lb x_{12}; Y =0\Rb \,\,=\,\,b^2_0\,\ln\Lb 1/\Lb x^2_{12}\,B\Rb\Rb;
\eeq
with $b^2_0 = 1/B$.

It should be stressed that \eq{IC} follows from the calculation of the amplitude for dipole- dipole scattering calculated in the Born Approximation of perturbative QCD. Introducing cutoff $B$ we still consider the same initial conditions, that follows from the perturbative QCD calculation, since our main goal to study the influence of the evolution in Y with the modified kernel of \eq{MODKER} on the behaviour of the scattering amplitude at large $b$.

It should be stressed that using the initial conditions of \eq{IC1} one can see that introducing $ \tilde{x}^2_{i k } \,=\,B\,x^2_{i k}$ the modified kernel
takes the form
\beq \label{STANDFORM}
 K^{B}\Lb x_{13},x_{32}|x_{12}\Rb\,\,\longrightarrow\,\,K^{1}\Lb \tilde{ x}_{13},\tilde{x}_{32}|\tilde{x}_{12}\Rb
 \eeq 
 while the initial conditions of \eq{IC1} can be re-written as $ {\cal N}^{BFKL}\Lb \tilde{x}_{12}; Y =0\Rb\,\,=\,\,\ln\Lb 1/\Lb \tilde{x}^2_{12}\Rb\Rb$.
 
 Therefore, in $\tilde{x}_{i k}$ the equation and the initial conditions have the same form for any values of $B$. However, we prefer to solve equations with the kernel $  K^{B}\Lb x_{13},x_{32}|x_{12}\Rb$ using the $B$ independence of the solution as the way to check the accuracy of our calculations.
 For numerical solution we again introduce the short distance regulator in the same way as in \eq{SOLNUM1} replacing $ K^{B}\Lb x_{13},x_{32}|x_{12}\Rb$ by the following expression
 \bea\label{SOLNUM2}
&&\int d^2 x_{13} K^{B}_{R}\Lb x_{13}, x_{32}| x_{12}\Rb\,{\cal N}^{BFKL}\Lb x_{12}; Y \Rb\,\,\equiv\,\,\\
  &&~~~~~~~~~~~\int d^2 x_{13} \, \frac{e^{ - B \Lb x^2_{13}\,+\,x^2_{32} \Rb}}{x^2_{32} + R^2}\,\Big\{ 2\,{\cal N}^{BFKL}\Lb x_{13}; Y \Rb  \,\,-\,\,2\,\frac{x^2_{12}}{x^2_{13}\,+\,x^2_{23}\,+\,2\,R^2}{\cal N}^{BFKL}\Lb x_{12}; Y \Rb\Big\}\nn
\eea
  
\subsection{Pomeron intercept}

\subsubsection{Numerical solution for the Pomeron inrtercept}
\begin{figure}
    \begin{minipage}{11cm}{
  \leavevmode
      \includegraphics[width=10cm]{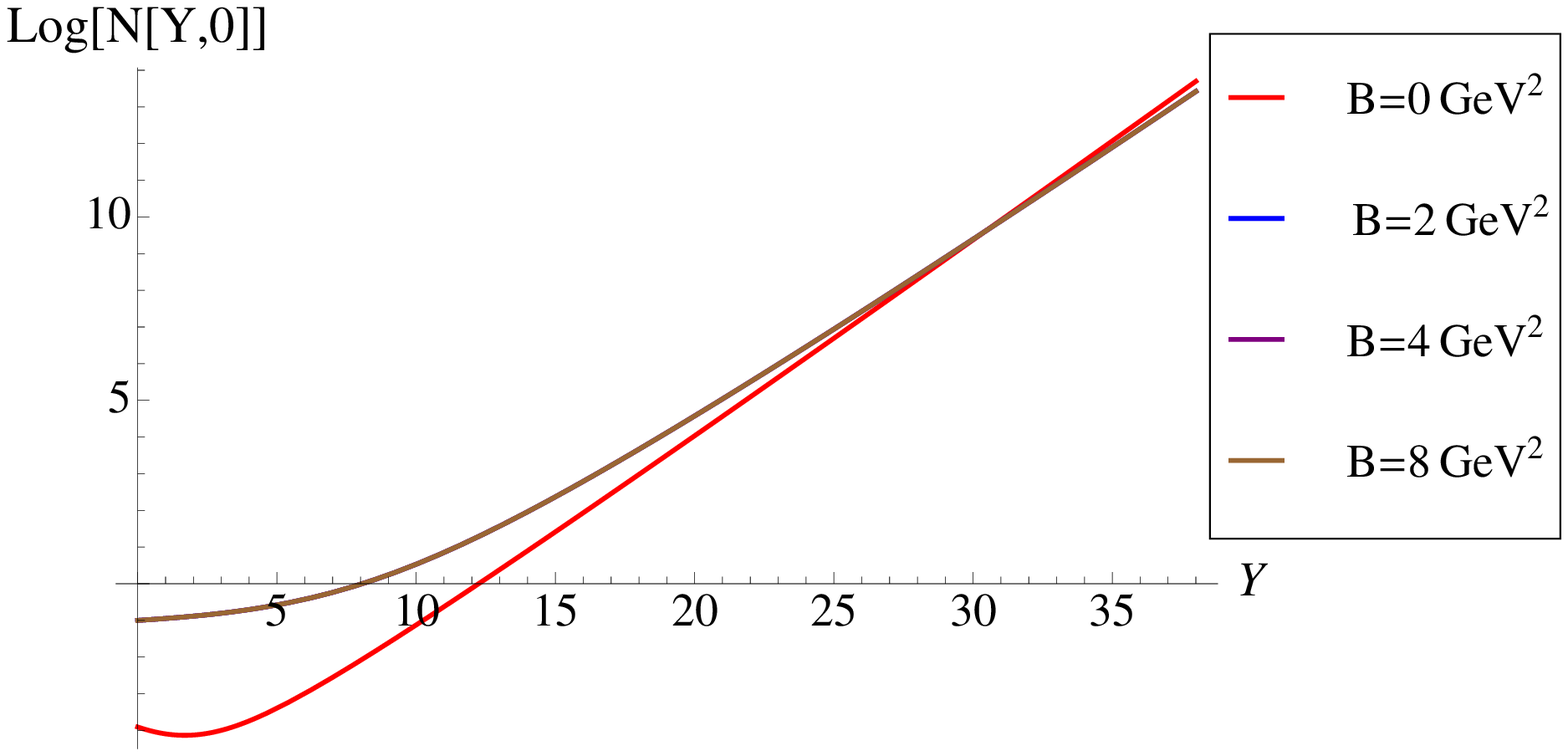}  }
      \end{minipage}
      \begin{minipage}{6cm}{
      \caption{The solutions to \protect\eq{EQ1} with $K^{B}_{R}\Lb x_{13}, x_{32}| x_{12}\Rb$  for  ${\cal N}\Lb x_{12}; Y \Rb$  as function of $Y$. $\bas$ is chosen to be equal to 0.2. $l = \ln\Lb x^2_{12}\,B\Rb = 0$. For the equation with the BFKL kernel ($B= 0$ in the figure) $l = \ln\Lb x^2_{12}\,\Lambda_{QCD}^2\Rb = 0$}
\label{pomint}}
\end{minipage}
   \end{figure}

\begin{figure}[ht]
    \begin{tabular}{ c c}
  \leavevmode
      \includegraphics[width=9cm]{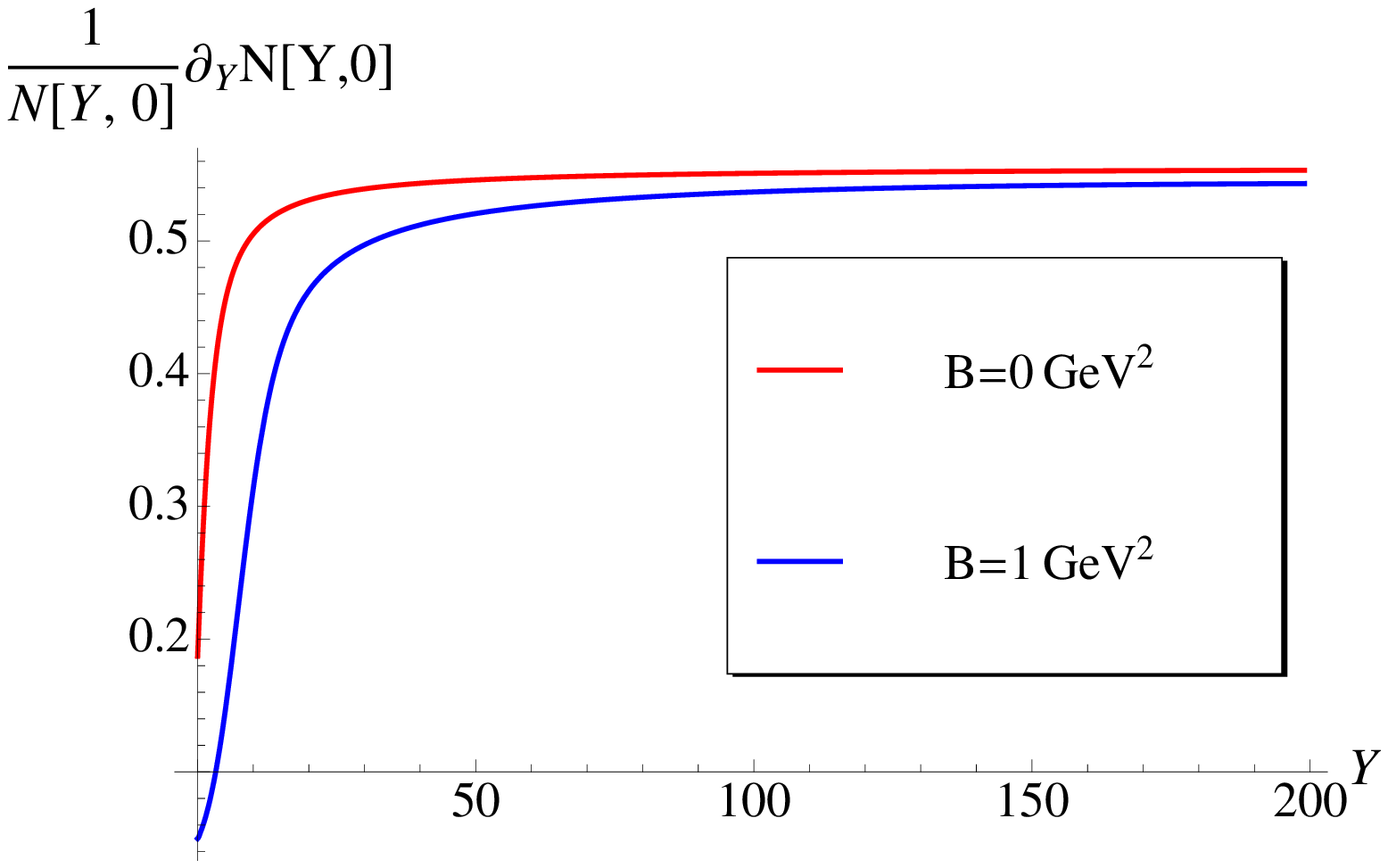} &  \includegraphics[width=9cm]{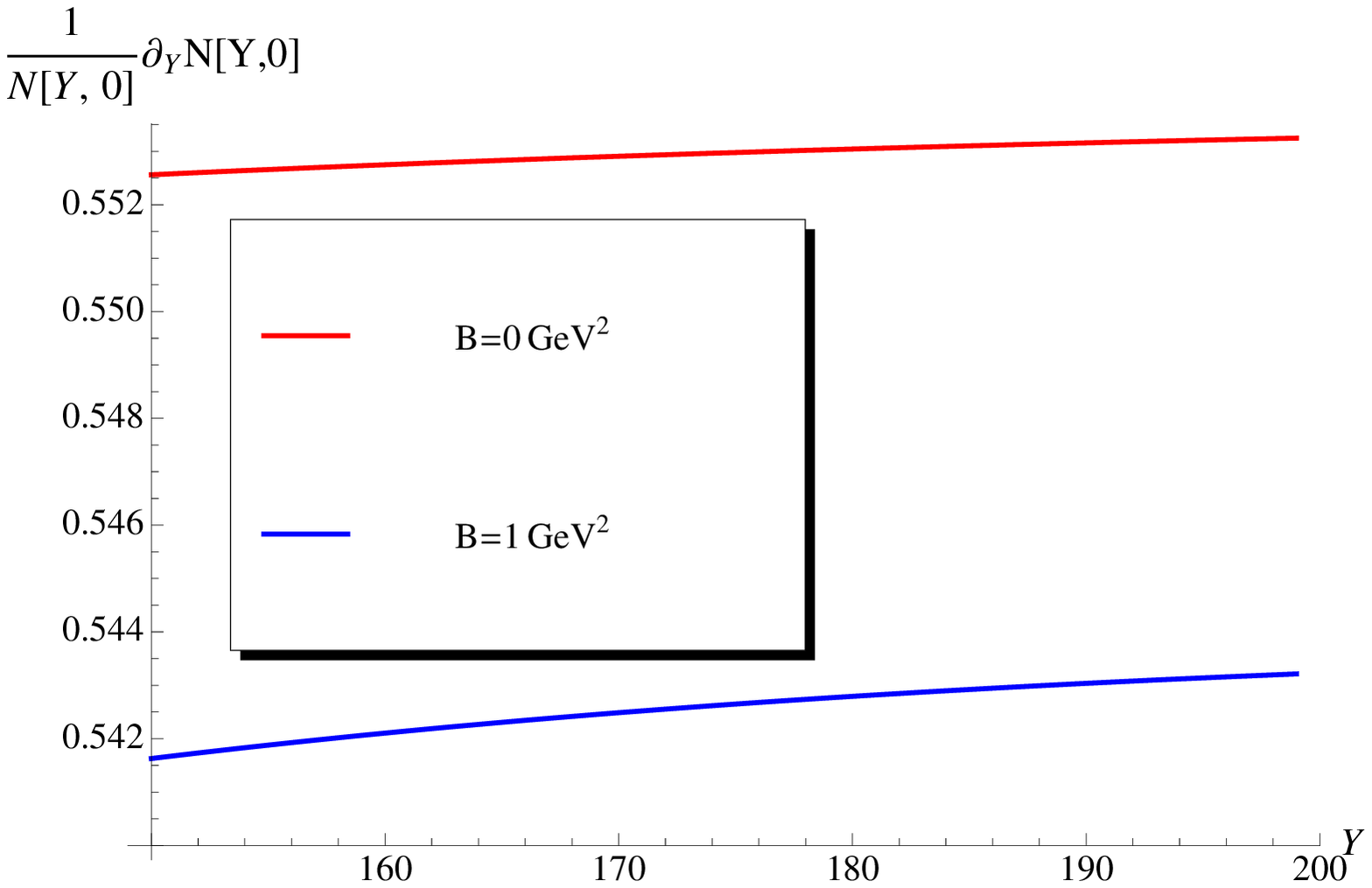}    \\
      \fig{pomint1}-a &\fig{pomint1}-b\\
         \end{tabular}
   \caption{  $d\,\ln{\cal N}\Lb x_{12}; Y \Rb\big{/} dY$  for the solutions to \protect\eq{EQ1} with $K^{B=1}_{R}\Lb x_{13}, x_{32}| x_{12}\Rb$ as function of $Y$. $\bas$ is chosen to be equal to 0.2. $l = \ln\Lb x^2_{12}\,B\Rb = 0$. For the equation with the BFKL kernel ($B= 0$ in the figure) $l = \ln\Lb x^2_{12}\,\Lambda_{QCD}^2\Rb = 0$}
\label{pomint1}
   \end{figure}

\begin{figure}
    \begin{minipage}{11cm}{
  \leavevmode
      \includegraphics[width=10cm]{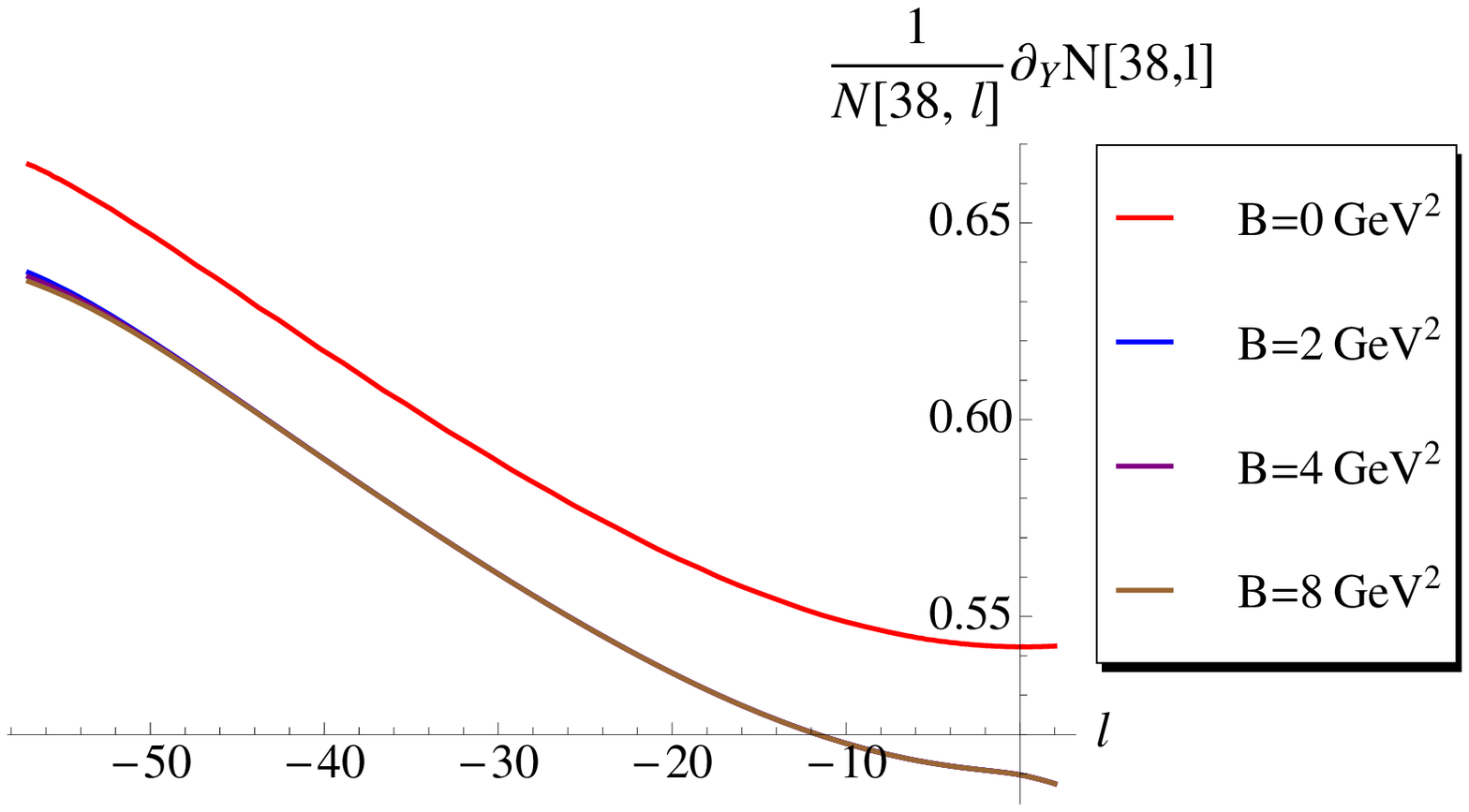}  }
      \end{minipage}
      \begin{minipage}{6cm}{
      \caption{$d\,\ln{\cal N}\Lb x_{12}; Y \Rb\big{/} dY$ for the solution  \protect\eq{EQ1} with $K^{B}_{R}\Lb x_{13}, x_{32}| x_{12}\Rb$    as function of $l$. $\bas$ is chosen to be equal to 0.2. $l = \ln\Lb x^2_{12}\,B\Rb $. For the equation with the BFKL kernel ($B= 0$ in the figure) $l = \ln\Lb x^2_{12}\,\Lambda_{QCD}^2\Rb = 0$. $Y = 38$.}
\label{pomint2}}
\end{minipage}
   \end{figure}

In \fig{pomint} it shown the solution to \eq{EQ1} with $K^{B}_{R}\Lb x_{13}, x_{32}| x_{12}\Rb$  for different values of $B$. One can see that   ${\cal N}\Lb x_{12}; Y \Rb$  grows as $\exp\Lb \om_0\,Y\Rb$ and the solution does not depend on the value of $B$. In this figure we do not show the dependence on the value of the regulator $R$ but  we actually studied this dependence in the same way as for solution to the BFKL equation and saw that the solution does not depend on the value of $R$ .  To see the power-like dependence of the solution in  a clearer way we plot in \fig{pomint1} $d \ln \Lb 
{\cal N}\Lb x_{12}; Y \Rb\Rb/d Y $ . We see that the value of the intercept is smaller than the BFKL intercept  but it is still increasing approaching this value. In \fig{pomint2} we show that $l$ dependence of the intercept is very similar to the BFKL Pomeron. 

\subsubsection{ Variational method}
The numerical solution suggests that the intercept of the Pomeron at $ B \neq 0$ is the same as for the BFKL Pomeron.
This fact as well as the influence of cutoff $B$  on $\om_0$  can be understood from the BFKL kernel in $\ga$-representation (see \eq{OMEGA}).  Indeed, this kernel appears in the calculation as the following integral ( see Refs.\cite{BFKL,LIREV,REV})
\beq \label{KERGA}
\chi\Lb \ga,0\Rb\,\,=\, \, \int\limits_0^1 dt \, \frac{t^{\gamma - 1}}{1-t} + \int\limits_1^\infty dt  
 \frac{t^{\gamma- 1} }{t-1} - \int\limits_0^\infty \frac{1}{t}
  \, \left[ \frac{1}{|t-1|} - \frac{1}{\sqrt{4 \, t^2 + 1}} \right]
\eeq

where $t \,=\,x^2_{13}/x^2_{12}$.

The first  term describes the increase of the dipole size due to evolution while the second one corresponds to normal DGLAP evolution in which the dipoles sizes decreases with the growth of $Y$. Introducing the modified BFKL kernel we cut the sizes of the intermediate dipoles such that $\tilde{x}^2_{13} \,\leq\,1$ ($t \leq 1/\tilde{ x}^2_{12}$.  Assuming $\tilde{x}^2_{12} \,\leq \,1$ we see that  we do not change integration over $t \,<\,1$ but $  t$ should be smaller than $t\, \leq \,1/\tilde{x}^2_{12}$. Therefore, we can estimate the modified kernel using
\beq \label{KERGA1}
\chi\Lb \ga,0; \tilde{x}_{12}\Rb \,\,=\, \, \int\limits_0^1 dt \, \frac{t^{\gamma - 1}}{1-t} + \int\limits_1^{1/\tilde{x}^2_{12}} dt  
 \frac{t^{\gamma- 1} }{t-1} - \int_0^{
 1/\tilde{x}^2_{12}} \frac{1}{t}
  \, \left[ \frac{1}{|t-1|} - \frac{1}{\sqrt{4 \, t^2 + 1}} \right]
\eeq
Introducing the variable $t = \tilde{x}^2_{12}/\tilde{x}^2_{12} $ which is smaller that 1,  we can re-write \eq{KERGA1} in the form
\beq \label{KERGA11}
\chi\Lb \ga,0; \tilde{x}_{12}\Rb \,\,=\, \, \int\limits_0^1 dt \, \frac{t^{\gamma - 1}\,-\,1}{1\,-\,t} + \int\limits_{\tilde{x}^2_{12}}^1  dt  
 \frac{t^{-\gamma}\,-\,1 }{1\,-\,t} \,+\, \int^\infty_{1/\tilde{x}^2_{12}} \frac{d t }{t\,\sqrt{4 \, t^2 + 1}} 
 \eeq

Taking the integral we obtain
\bea \label{KERGA2}
\chi\Lb \ga,0; \tilde{x}_{12}\Rb \,\,&=&\,\, \chi^{BFKL}\Lb \ga,0\Rb \,+\, \mbox{arccsch}\Lb 2/\tilde{x}^2_{12} \Rb \,-\,B\Lb \tilde{x}_{12}; 1 - \ga, 0\Rb  \,-\,\ln\Lb 1 - \tilde{x}^2_{12}\Rb
\eea
where $B\Lb \tilde{x}_{12}; 1 - \ga, 0\Rb$ is incomplete $B$ function and $\chi^{BFKL}\Lb \ga,0\Rb$ is given by \eq{OMEGA}.

One  can see that \eq{KERGA2} depends on $x^2_{12}$ through $\tilde{x}_{12}$ and, therefore,  $\om_0\Lb \ga,0,\tilde{x}_{12}\Rb\,=\,\bas\,\chi\Lb \ga,0; \tilde{x}_{12}\Rb $  cannot be the intercept of the Pomeron since the intercept cannot depend on the sizes of dipoles.
However, we can use \eq{KERGA2} for developing the estimate in the variational approach for finding the ground state (the maximal intercept).
Indeed, if we   introduce\footnote{In this section as well as in sections 3.1.3 and 3.1.4 we use variable ${\cal Y}$ which is equal to $\bas Y$(${\cal Y}\,=\,\bas Y$).}
\beq \label{BFKLWF}
{\cal N}^{BFKL}\Lb x_{12}; Y \Rb\,\,=\,\,\int \frac{d \om}{2 \pi i}\,e^{ \om {\cal Y}}\,\Psi\Lb \tilde{x}_{12},\om
\Rb
\eeq
the equation looks as 
\beq \label{STAT}
\om \Psi\Lb \tilde{x}_{12}, \om \Rb\,\,=\,\,\int d^2 \tilde{ x}_{13} \,K^{B}\Lb \tilde{x}_{13}, \tilde{x}_{23}| x_{12} \Rb \Psi\Lb\tilde{ x}_{13}, \om\Rb 
\eeq
where the r.h.s. of the equation has been discussed in \eq{SOLNUM2}.
For the BFKL equation the eigenfunction $  \Psi_{BFKL}\Lb x_{13}, \om\Rb =  \Lb x^2_{13}\Rb^{\ga - 1}$ and the eigenvalue is given by \eq{OMEGA}.
This function cannot be a eigenfunction of \eq{BFKLWF} at $ B \neq 0$ as we have discussed. However, we can use this BFKL eigenfunction as a trial function in the variational method for searching the maximal value of the intercept. Actually, we use as the trial function 
\beq \label{TRF}
\Psi_{tr}\Lb \tilde{x}_{12},\ga\Rb\,\,=\,\, \Psi_{BFKL}\Lb x_{12}, \ga \Rb\,\Theta\Lb 1 - \tilde{x}^2_{12}\Rb\,\,=\,\,\Lb \tilde{x}^2_{12}\Rb^{\ga - 1}\,\Theta\Lb 1 - \tilde{x}^2_{12}\Rb
\eeq
The  variational principle has the following form for the problem of finding the maximal intercept
\bea \label{VP}
\om_{max}\,\,\geq\,\,\om\Lb \ga \Rb \,\,&=&\,\,\frac{\int d^2 \tilde{x}_{12} \Psi ^*_{tr} \Lb \tilde{x}_{12},\ga\Rb \int d^2 \tilde{ x}_{13} \,K^{B}\Lb \tilde{x}_{13}, \tilde{x}_{23}| x_{12} \Rb \Psi_{tr}\Lb\tilde{ x}_{13}, \ga\Rb}{ \int d^2 \tilde{x}_{12} \,\Psi ^*_{tr} \Lb \tilde{x}_{12},\ga\Rb \,\Psi _{tr} \Lb \tilde{x}_{12},\ga\Rb} \nn\\
& &\nn\\
  & =& \frac{\int d^2\tilde{x}_{12}\,\Psi ^*_{tr} \Lb \tilde{x}_{12},\ga\Rb \,\chi\Lb \ga,0; \tilde{x}_{12}\Rb \Psi _{tr} \Lb \tilde{x}_{12},\ga\Rb}{ \int d^2 \tilde{x}_{12} \,\Psi ^*_{tr} \Lb \tilde{x}_{12},\ga\Rb \,\Psi _{tr} \Lb \tilde{x}_{12},\ga\Rb}
  \eea
  From convergency of  $\int d^2 \tilde{x}_{12} \,\Psi ^*_{tr} \Lb \tilde{x}_{12},\ga\Rb \,\Psi _{tr} \Lb \tilde{x}_{12},\ga\Rb$ we conclude that $ \ga \geq 0.5$. 
  
     \begin{figure}[ht]
    \begin{minipage}{10cm}{
  \leavevmode
      \includegraphics[width=9cm]{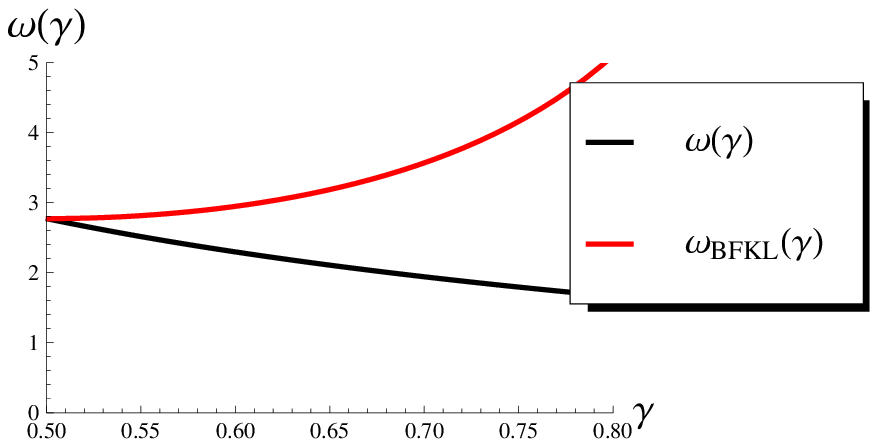}  }
      \end{minipage}
      \begin{minipage}{6cm}{
      \caption{$\om\Lb \ga\Rb$ (see \protect\eq{VP})  and $
     \ om_{BFKL}\Lb \ga\Rb$(see \protect\eq{OMEGA}) versus $\ga$( black and red lines, respectively).  In this figure we redefine $\om$ dividing it by $\bas$. }
\label{vp}}
\end{minipage}
   \end{figure}

 From \fig{vp} we see that $\om(\ga=1/2)$ coincide with the BFKL value. In other words, we prove that the resulting maximal $\om$ can be either equal to the BFKL one or larger than the  BFKL value. 
\subsubsection{Semi-classical solution}

Examining the property of the solution to the modified BFKL equation we
    wish to find a semi-classical solution to this equation searching it in the form:
  \beq\label{SC1}
 N\Lb {\cal Y}; l\Rb\,\,=\,\,e^{S\Lb {\cal  Y}, l \Rb}\,\,=\,\,e^{\om\Lb {\cal Y}, l\Rb Y \,\,+\,\,\Lb \ga\Lb {\cal Y}, l\Rb - 1|\Rb l} ~ \,
  \mbox{where}~~\om\Lb {\cal Y}, l\Rb\,=\,\frac{\D S\Lb {\cal Y};  l\Rb}{\D {\cal Y}};~~\ga\Lb {\cal Y},l\Rb\,-\,1\,=\,\frac{\D S|\Lb {\cal Y}; l\Rb}{\D l} \eeq
 with smooth functions  $\om\Lb {\cal Y}, l\Rb$ and    $\ga\Lb {\cal Y}, l\Rb$.
 
  Inserting \eq{SC1} to the equation we obtain:
 \beq \label{EQSC}
 \om\Lb {\cal Y}, l\Rb  \,-\, \chi\Lb\ga, 0, x^2_{12}\Rb  \,\,=\,\,0
 \eeq
 Deriving \eq{EQSC} we use that $\ga$ is a smooth function and performing the integral of \eq{KERGA11} we can consider $\ga$ as being a constant.  Since the equation is the first order differential equation in respect to $Y$ the condition  for applying the semi-classical approach looks as follows
 \beq \label{SC2}
 \frac{\partial \ga \Lb {\cal Y}; l \Rb/\partial l}{\Lb 1 -  \ga \Lb {\cal Y},  l \Rb\Rb^2}\,\,\ll\,\,1;~~~~~~~~~~ l\, \frac{\partial \ga \Lb {\cal Y}; l \Rb}{\partial l}\,\,\ll\,\,1
  \eeq
 where $l \,=\,\ln\Lb x^2_{12}/\Lambda_{QCD}\Rb$.

   It is known\cite{MATH} 
that for the equation in the form
\beq \label{SC3}
F\Lb {\cal Y}, l, S,\,\gamma,\omega\Rb\,\,=\,\,0
\eeq
we can introduce the set
of
characteristic lines :  $l(t), {\cal Y}(t), S(t)$, $ \omega(t)$ and $
\gamma(t)$ which are the
functions of the variable $t$ ( artificial time), that satisfy the following
equations:
\begin{eqnarray}
&&\hspace{-0.3cm}(1.)\,\,\,\,\frac{d l}{d\,t}\,\,=\,\,F_{\gamma}\,\,= \,\,- \frac{d \chi\Lb \ga, 0, l \Rb}{d \ga} \nn\\
&&\hspace{-0.3cm}(2.)\,\,\,\,\,\,\,\label{SC4}
\frac{d{\cal Y}\Lb t\Rb}{d\,t}\,\,=\,\,F_{\omega}\,\,=\,\,1\,\,\,\,\,\,\,\nn\\
 &&\hspace{-0.3cm}(3.)\,\,\,\,
\frac{d S}{d\,t}\,\,=\,\,\gamma\,F_{\gamma}\,+\,\omega\,F_{\omega}\,\,=\,\,  \,\,-\Lb \ga - 1\Rb\frac{\partial \chi\Lb \ga, 0, l \Rb}{\partial \ga} \,\,+\,\,\om
\nonumber \\
&&\hspace{-0.3cm}(4.)\,\,\,\,\frac{d\,\gamma}{d\,t}\,\,=\,\,-
(\,F_{l}\,+\,\gamma\,F_{S}\,)\,\,=\,\,  \frac{\D  \chi\Lb \ga(t), 0, l(t) \Rb}{\D\, l }
\end{eqnarray}

  The fifth equation is \eq{EQSC}.

First, we see two thing which simplify a bit the equation: we can consider $t = {\cal Y}$ and dividing \eq{SC4}- 4 by \eq{SC4}-1 we see that
\beq \label{SC5}
\frac{d\,\gamma}{d\,l}\,\,=\,\,- \frac{ \frac{\D \chi\Lb\ga(t), 0,l(t)\Rb}{\D\, l}}{ \frac{d \chi\Lb \ga, 0, l(t) \Rb}{d \ga} }\,\,
\eeq

We solve this equation putting $\ga\Lb l = \infty\Rb$ which coincide with the solution to the BFKL ,  as the initial condition. Due to convergency  of the integral for the norm of $N$ we know that $\ga\Lb \infty\Rb \,\geq\,0.5$.

The second step after finding $\ga\Lb l \Rb$ is to solve \eq{SC4}-1 to find the form of trajectories. It tuns out that we do not need to find $l\Lb {\cal Y}\Rb$ as a function of ${\cal Y}$ for finding the values of $\omega$.

\begin{figure}[ht]
     \begin{center}
     \includegraphics[width=10cm]{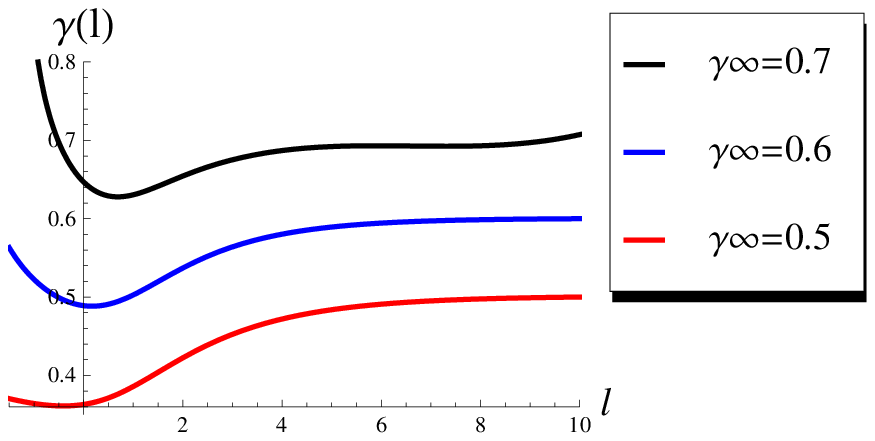} 
     \end{center}    
      \caption{ The trajectories of \protect\eq{EQSC} at different values of $\ga\Lb \infty\Rb$  }
\label{sc1}
   \end{figure}

\begin{figure}[ht]
     \begin{center}
     \includegraphics[width=10cm]{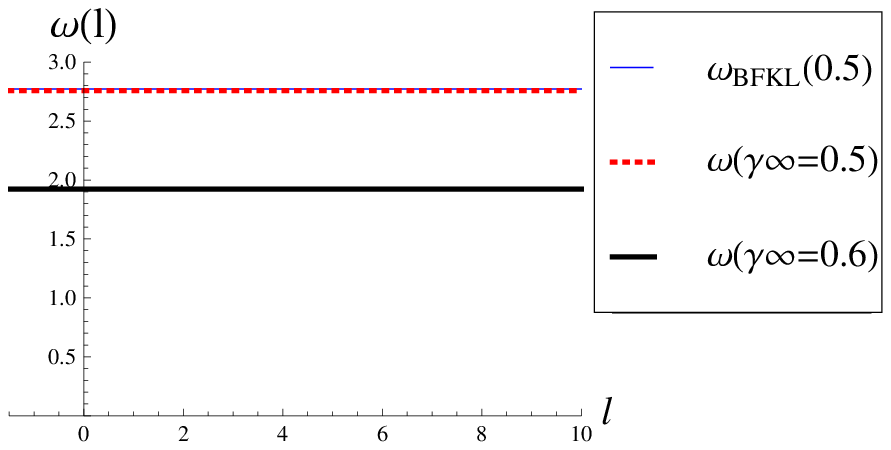} 
     \end{center}    
      \caption{ $\om\Lb  l \Rb $ versus $l$ at different values of $\ga\Lb \infty\Rb$. Notice that  $\om_{\mbox{\tiny BFKL}}\Lb 0.5\Rb$ (blue line) and $\om\Lb \ga_{\infty} = 0.5\Rb$ (red dotted line) are the same. }
\label{sc2}
   \end{figure}

The last the third step is to find from \eq{EQSC} the value of $\om$.
In \fig{sc1} we plot function $\ga\Lb l \Rb$ for different $\ga_\infty\,=\,\ga\Lb l = \infty\Rb$, while \fig{sc2} shows the dependence of $\omega\Lb \ga \Rb \,=\, \om\Lb {\cal Y}, l\Rb  \,-\, \chi\Lb\ga, 0, l \Rb $ versus $l$.

   From \fig{sc2} one can see that the values of $\om\Lb\ga\Lb l \Rb\Rb$ does not depend on $l$ and these values turn out to be smaller  or equal to the values of the intercept for the BFKL equation.
   
\begin{figure}[ht]
     \begin{center}
     \begin{tabular}{c c c}
     \includegraphics[width=7cm]{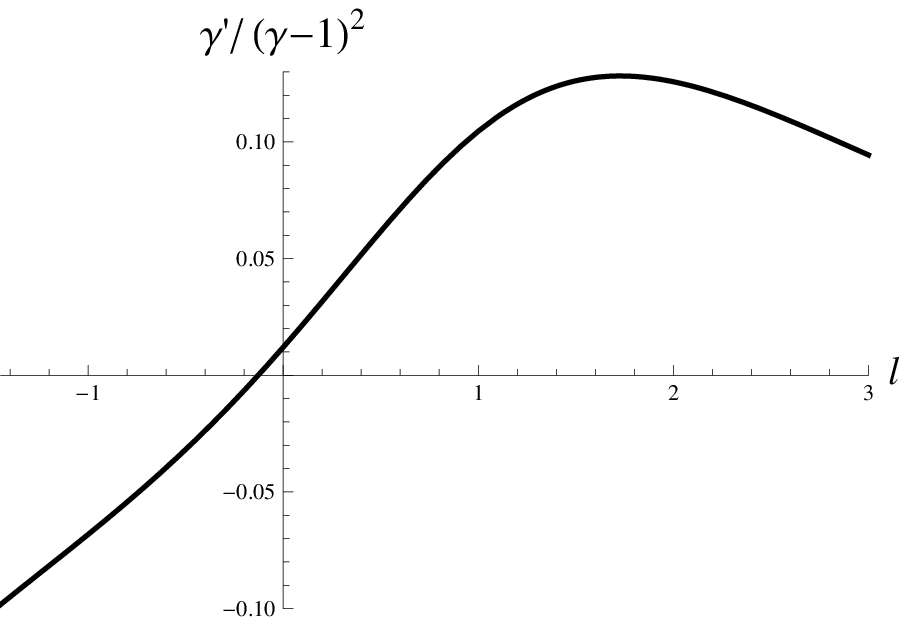} &~~~~~~~~& \includegraphics[width=7cm]{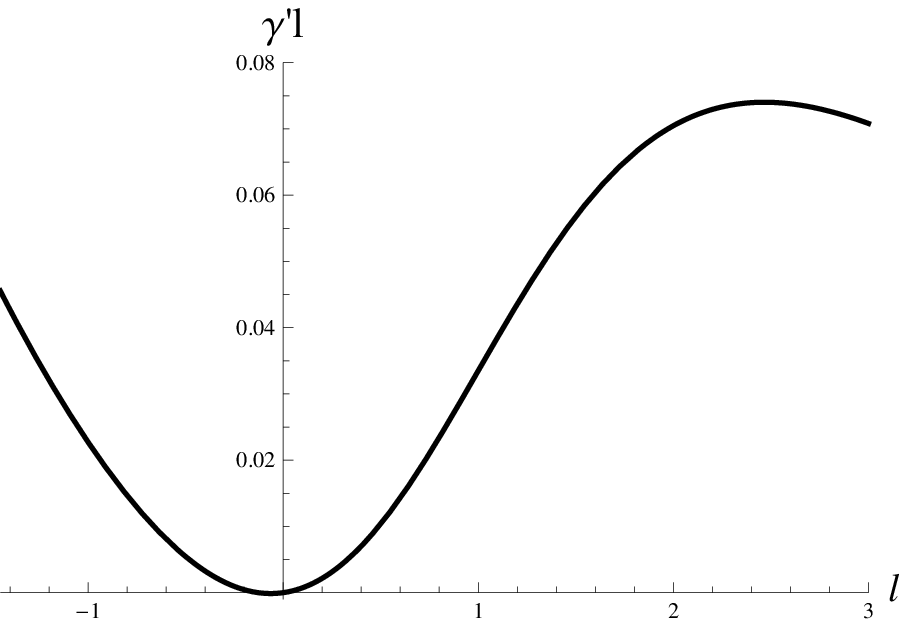}  \\
     \fig{sc3}-a &      & \fig{sc3}-b\\
      \end{tabular}   
      \end{center} 
      \caption{ The ratio $d \ga\Lb l \Rb/d l\Big{/}\Lb 1 - \ga\Lb l \Rb\Rb^2$ (see \protect\fig{sc3}-a) and the product $l \,d \ga\Lb l \Rb/d l   $ (see \protect\fig{sc3}-b)   versus $l$  for $\ga\Lb \infty\Rb = 0.5$.  }
\label{sc3}
   \end{figure}

In \fig{sc3} we plot the ratio $d \ga\Lb l \Rb/d l\Big{/}\Lb 1 - \ga\Lb l\Rb\Rb^2$ and the product $ l d \ga\Lb l \Rb/d l$ as  a function of $l$.  One can see that  both of these observables  turn out to be small and, therefore, we can trust the semi-classical approach.

 Concluding this subsection we see that the semi-classical approach leads to $\om \,=\,\om_{BFKL}$.
 
\subsubsection{Diffusion approximation}

In direct analogy with the BFKL equation we develop in this section the diffusion approximation to the modified BFKL equation. The main idea of this approximation is to introduce a new function
\beq \label{NBAR}
\bar{N}\Lb {\cal Y}, l\Rb\,\,=\,\,e^{\h\, l  }N\Lb {\cal Y}, l\Rb\,\,=\,\,\int^{\epsilon + i \infty}_{\epsilon - i \infty}\frac{d \omega}{ 2 \pi i}e^{\omega \,{\cal Y}}\,\bar{n}\Lb \omega,l\Rb\,\,=\,\,\int^{\epsilon + i \infty}_{\epsilon - i \infty}\frac{d \omega}{ 2 \pi i} \int^{i\epsilon +  \infty}_{i\epsilon - \infty} \frac{d \nu}{ 2 \pi i}\bar{n}\Lb \om, \nu\Rb\,e^{\omega \,{\cal Y} \,+\, i \nu l}
\eeq
The diffusion approximation  means that we can reduce the modified BFKL equation to the differential equation  using the following expansion for $ \bar{n}\Lb \omega,l\Rb$:
\beq \label{SERN}
\bar{n}\Lb \omega,l'\Rb\,\,=\,\,\bar{n}\Lb \omega,l\Rb\,\,+\,\,\frac{\partial \bar{n}\Lb \omega,l'\Rb}{\partial l'}|_{l'=l}\Lb l' - l \Rb\,\,+\,\,\h \frac{\partial^2 \bar{n}\Lb \omega,l'\Rb}{\partial l'^2}|_{l' = l} \Lb l - l\Rb^2\,+\,\dots
\eeq
The equation takes the form after plugging in \eq{SERN} into \eq{STAT}
\beq \label{EQDIFF}
\Lb \om\,-\,\Delta\Lb l \Rb \Rb\bar{n}\Lb \omega,l\Rb\,\,-\,d1\Lb l \Rb \,\frac{\partial \bar{n}\Lb \omega,l\Rb}{\partial l}\,\,-\,\,\h d2\Lb l \Rb\, \frac{\partial^2 \bar{n}\Lb \omega,l\Rb}{\partial l^2} \,\,=\,\,0
\eeq

Functions $\Delta\Lb l \Rb$,$d1\Lb l \Rb$\,  and $d2\Lb l \Rb$ can be expressed through the kernel $\chi\Lb \h + i \nu,0,e^{\h l}\Rb$ that has been introduced in \eq{KERGA2}, viz,
\beq
\Delta\Lb l \Rb =\chi\Lb \h,0,e^{\h l}\Rb;~~~d1\Lb l \Rb =-i \frac{\partial \chi\Lb \h + i \nu,0,e^{\h l}\Rb}{\partial \nu}|_{\nu = 0};~~~
d2\Lb l \Rb=- \frac{\partial^2\chi\Lb \h + i \nu,0,e^{\h l}\Rb}{\partial \nu}^2|_{\nu = 0}.
\eeq

\begin{figure}[h]
   \begin{tabular}{c c c}
    \includegraphics[width=5cm]{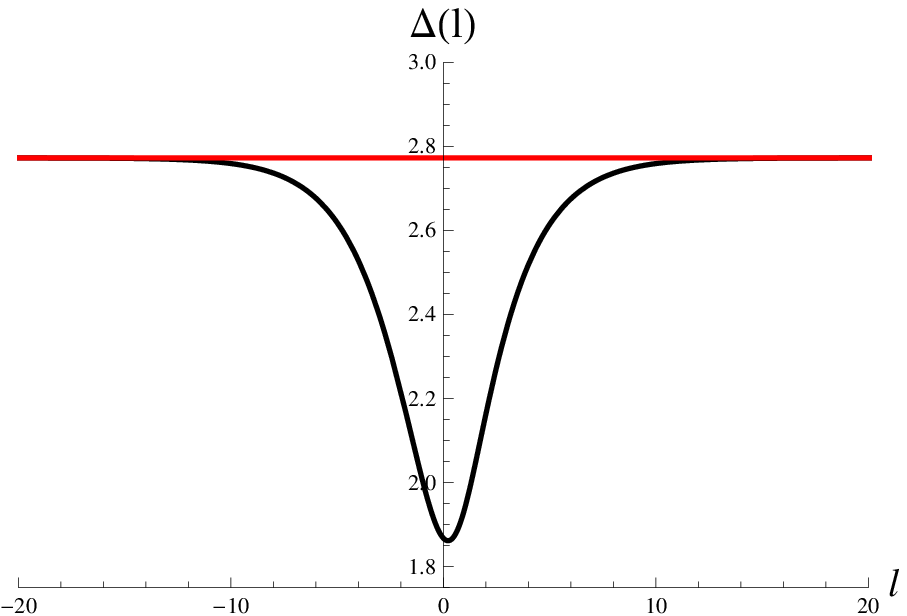} &    \includegraphics[width=5cm]{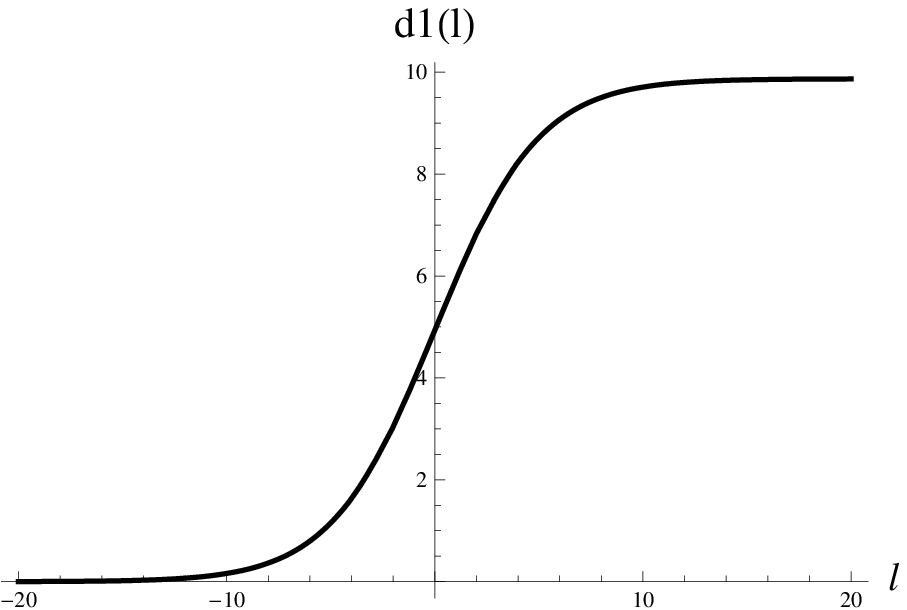}   &  \includegraphics[width=5cm]{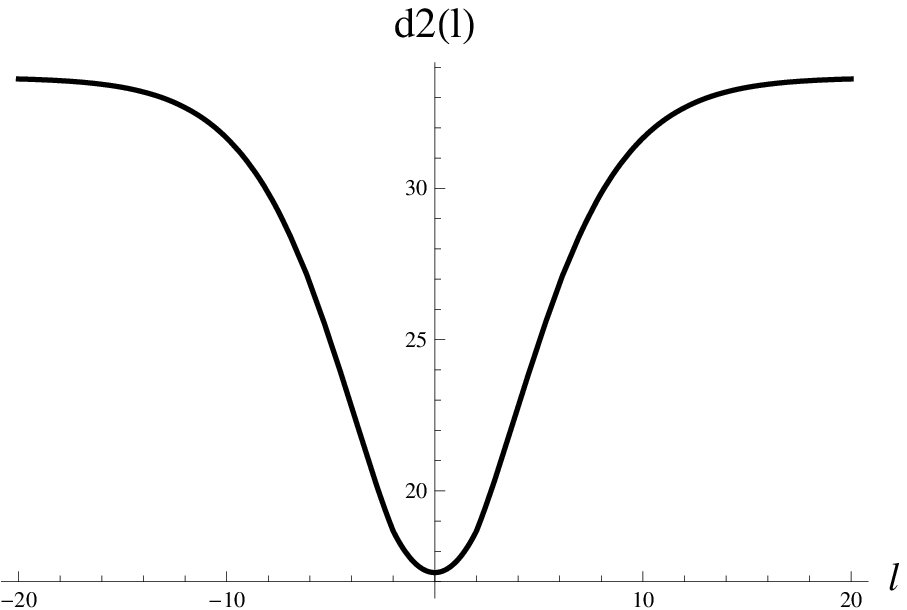}\\
    \fig{fun}-a & \fig{fun}-b &\fig{fun}-c\\
    \end{tabular}
    \caption{Functions $\Delta\Lb l \Rb$, $d1\Lb l \Rb$ and $d2\Lb l\Rb$ versus $l$. The red line shows the BFKL value for $\om_{\mbox{\tiny BFKL}}\Lb l \Rb\,\,=\,\,4\ln 2$.}    \label{fun}
    \end{figure}

   \fig{fun} shows these functions.

     Introducing $ \bar{n}\Lb \omega,l\Rb \,\,=\,\,\exp\Lb \phi\Lb \om, l \Rb\Rb$ for function $\phi$ \eq{EQDIFF} reduces to
     \beq \label{EQDIFFPH}
\Lb \om\,-\,\Delta\Lb l \Rb \Rb\,\,=\,\,d1\Lb l \Rb \ga \Lb \om, l\Rb\,\,+\,\,\h d2\Lb l \Rb\, \Lb \frac{\partial \ga\Lb \omega,l\Rb}{\partial l}\,+\,\ga^2\Lb \om, l\Rb\Rb 
\eeq
where $\ga \,=\,\frac{\partial \phi\Lb \om, l \Rb}{\partial l}$.  
     
     For $l \to -\infty$   $\Delta\Lb l \Rb \to \om_{\mbox{\tiny BFKL}}  $, $d1\Lb l \Rb \to 0 $ and $ d2\Lb l \Rb \,\to\,D_0 $(see \eq{DIFF}).
     
     Therefore, for large and negative $l$ the solution for
     \beq \label{INC}
      \ga\Lb \om , l\Rb \,\xrightarrow{|l|\,\gg\, 1}\,  \sqrt{\Lb \om - \om_{\mbox{\tiny BFKL}}\Rb/(2 D_0)} .
      \eeq
       These values give us the initial condition for \eq{EQDIFFPH}.

\begin{figure}[h]
   \begin{tabular}{ c c}
    \includegraphics[width=8cm]{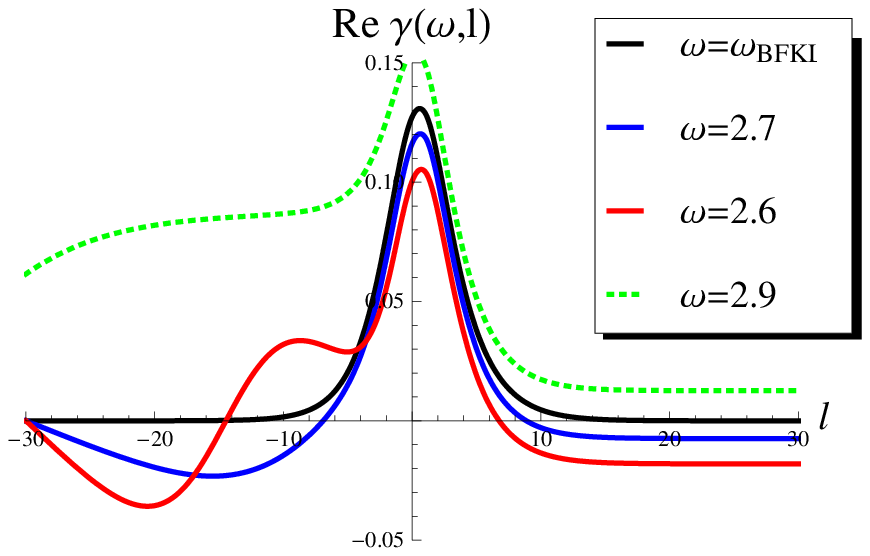} &    \includegraphics[width=6.8cm]{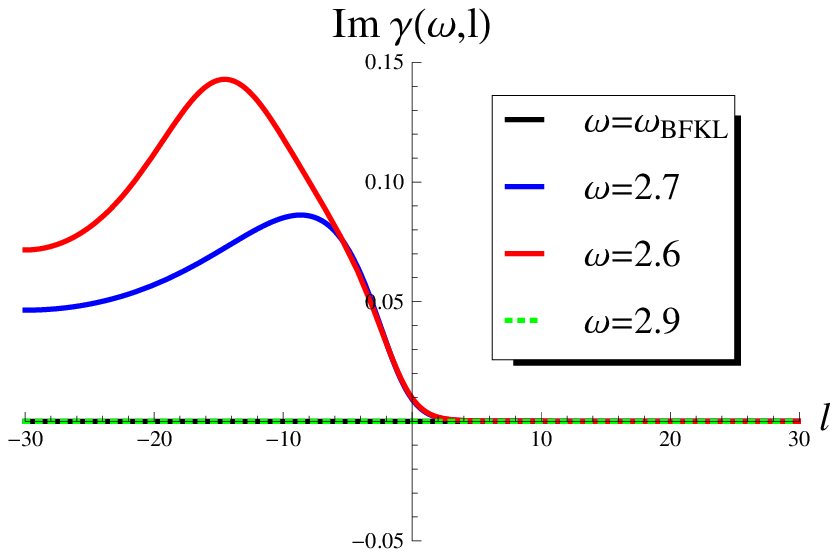} \\
    \fig{ga}-a & \fig{ga}-b    \\
        \end{tabular}
    \caption{Solution to \protect\eq{EQDIFFPH} with the initial condition of \protect\eq{INC}  versus $l$: \protect\fig{ga}-a for $ \mbox{Re} \ga\Lb \om, l\Rb$ and \protect\fig{ga}-b for $ \mbox{Im} \ga\Lb \om, l \Rb$. }
    \label{ga}
    \end{figure}

     Solving \eq{EQDIFFPH} numerically we see that for all $\om \,\leq \,\om_{\mbox{\tiny BFKL}}$  $\ga \leq 0$ at $l \to +\infty$. Such $\ga$  lead to the amplitude $\bar{n}$ which normalization is convergent integral, namely,
     \beq \label{NORM}
   \int d l |\bar{n}\Lb\om, l\Rb|^2\,d l \,\,<\,\, \infty
   \eeq
   while for $\om \,>\,\om_{\mbox{\tiny BFKL}}$ $\ga$ at large $l > 0$ is positive and the integral in \eq{NORM} is divergent (see \fig{ga}).

   In other words, our spectrum of $\om$ is continuous with $\om \,\leq \,\om_{\mbox{\tiny BFKL}}$.

\subsection{Pomeron slope}


     \begin{figure}
    \begin{center}
  \leavevmode
      \includegraphics[width=11cm]{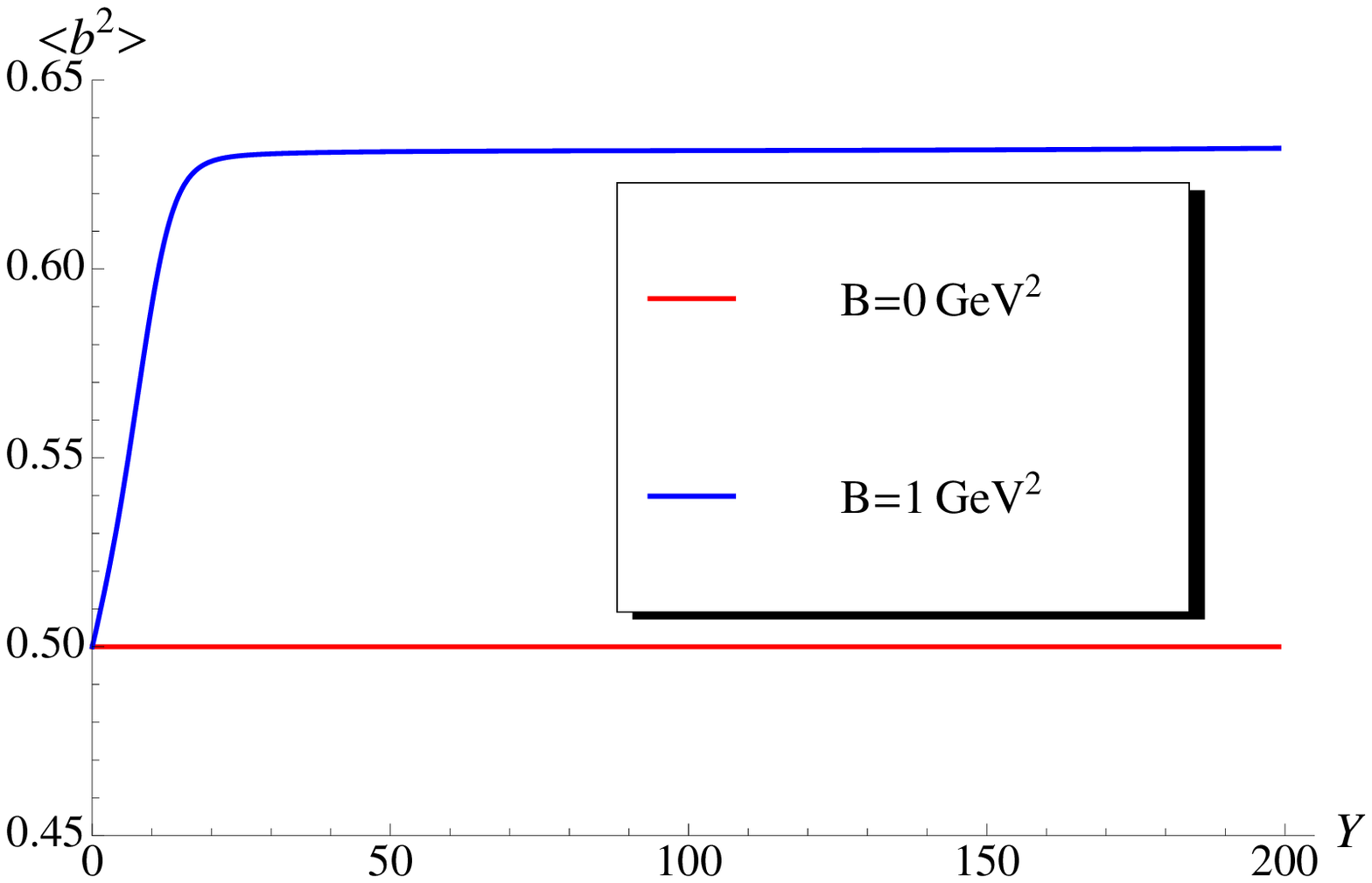}
      \end{center}
      \caption{ $\langle|b^2\Lb Y,l\Rb|\rangle$ versus $Y$ at different values of $B$ (see \protect\eq{MODKER}). $\bas = 0.2$. For both $B=0$ and $B=1 GeV^2$ we took $b_0$ in \protect\eq{ICBFKL} and \protect\eq{IC1} the same and  equal to $0.5 \,GeV^{-2}$.}
\label{pomslY}
   \end{figure}

Approaching the problem of calculation of $\langle|b^2|\rangle$, the first observation that we can make is that \eq{EQ2} is valid also for the modified BFKL kernel. Indeed,  the kernel itself does not depend on the impact parameter  and  functions $E^{n,\ga}\Lb \rho_{1 a},\rho_{2 a}\Rb$ is the complete set of functions. 
As we noticed in derivation of \eq{EQ2} the term in $\Big\{\dots\Big\}$ in \eq{BX} vanishes due to invariance of function  $E^{n=0,\ga}\Lb \rho_{1 a},\rho_{2 a}\Rb$
with respect to transformation $\vec{b} \,\to\,- \,\vec{b}$.  Since $\omega\Lb \ga, n\Rb \,\leq\,0$ for $n \,\geq \,1$ the only component of the arbitrary function of the initial condition that survives at large $Y$ is  its projection on $E^{n=0,\ga}\Lb \rho_{1 a},\rho_{2 a}\Rb$ . For this projection the term of \eq{BX} vanishes
leading to \eq{EQ2}.

\begin{figure}[th]
     \begin{center}
    \begin{tabular}{c c c}
  \leavevmode
      \includegraphics[width=6.5cm]{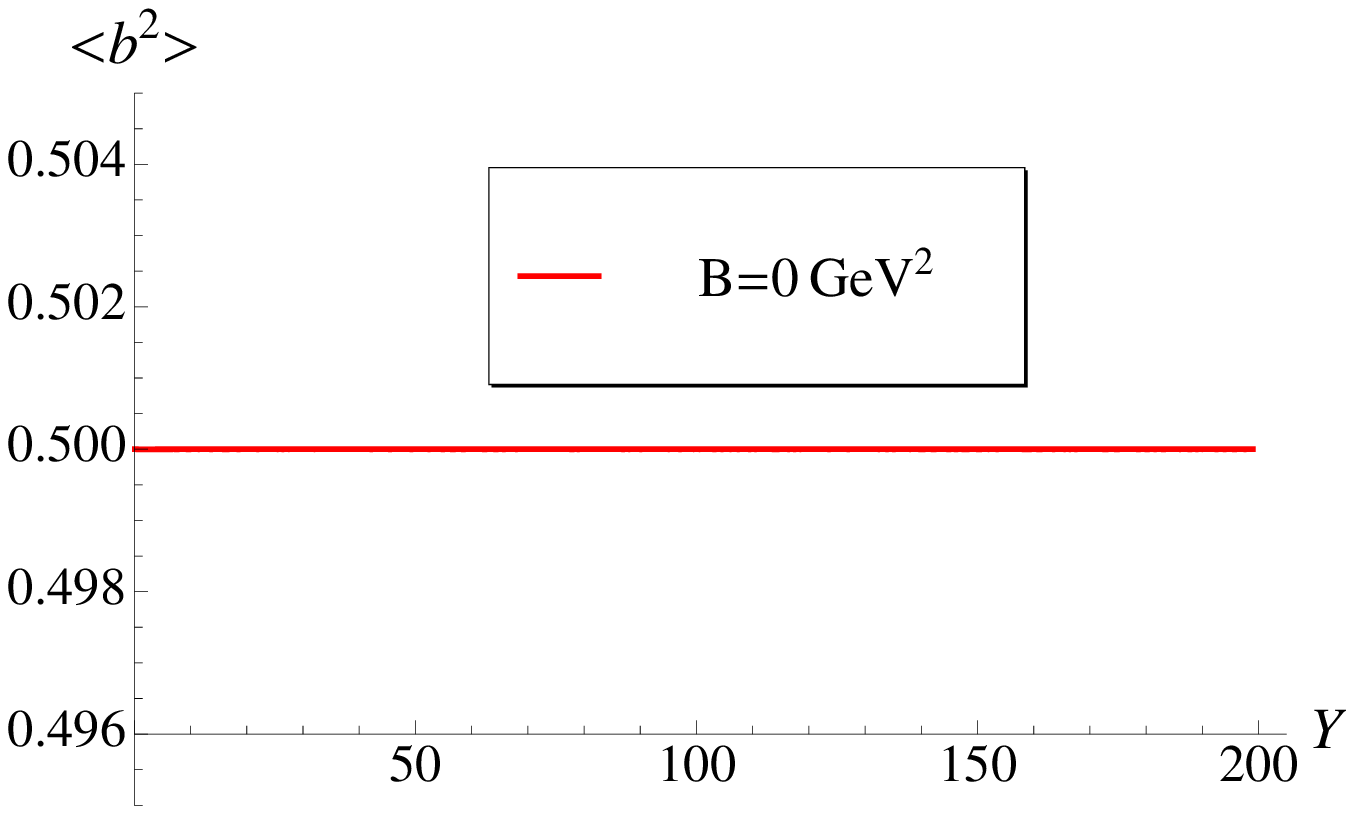}  &  ~~~~~~~~~~&\includegraphics[width=6.5cm]{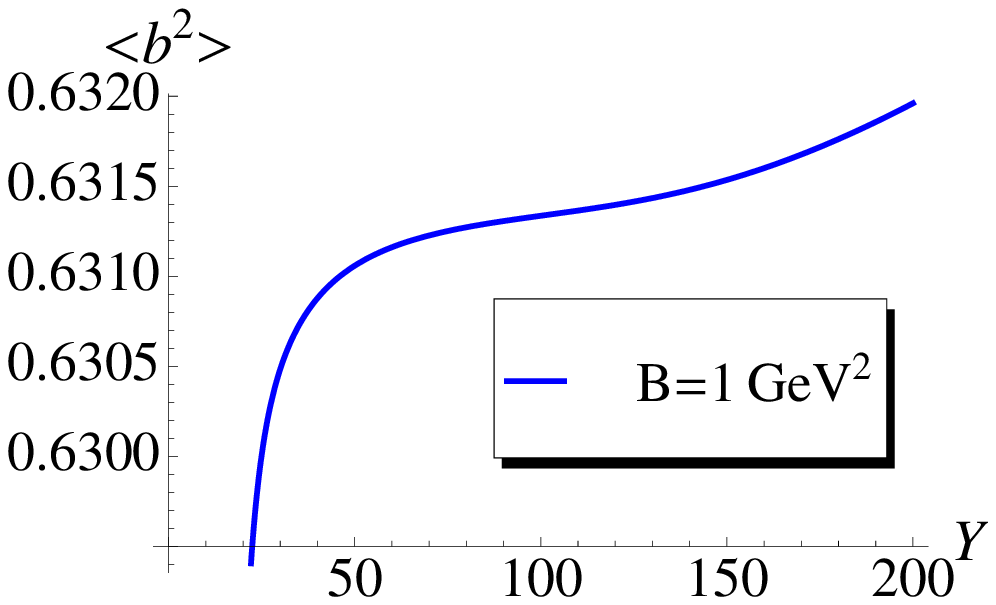}\\
      \end{tabular}
      \end{center}
\caption{$ \langle|b^2\Lb Y,l\Rb|\rangle= 4 \alpha'_\pom\, Y$ versus $Y$ at different values of $B$ (see \protect\eq{MODKER}). $\bas = 0.2$.}
\label{pomslYz}
   \end{figure}

Solving \eq{EQ2} with the kernel of \eq{MODKER} we calculate (see \eq{B2}) $\langle|b^2|\rangle$ as function of $Y$ (see \fig{pomslY}). In the Reggeon approach this value gives the information on the slope of the Pomeron trajectory ($\alpha'_\pom$) since
$\langle|b^2|\rangle\,\,\xrightarrow{ Y\,\gg\,1}\,\,4 \,\alpha'_\pom\,Y$.  One can see from \fig{pomslY} that $\langle|b^2|\rangle$ does not depend on $Y$ at large values of $Y$. Actually, the solution to the modified BFKL equations shows a weak $Y$ dependence (see \fig{pomslYz} which is zooomed \fig{pomslY}).
 However, even if we assume that $\langle|b^2|\rangle|_{B=1}\,=\,4 \alpha'_\pom Y$ the value of $\alpha'_\pom\,\, \leq\,\,0.5 10^{-5}\,b_0$ is extremely small.
 Hence we can conclude  that the modified BFKL Pomeron has $\alpha'_\pom\,\approx\,0$.
 

 This result was expected and its explanation based on the general features of QCD.  The general origin of  the increase of $\langle|b^2|\rangle $ with $Y$ (
 The relation: $\langle|b^2|\rangle\,=\,4 \alpha'_\pom Y$) ,   was understood in 70's by V.N.Gribov (Gribov's diffusion \cite{GRIB}). Each emission leads to change in impact parameter by $\Delta b^2 = 1/p^2_T$ where $p_T$ is the typical transverse momentum.  After $n$ emission $\langle b^2_n \rangle\, \,=\,\,\Delta b^2\,n$ which corresponds to the random walk in the transverse plane. Since the number of emission is proportional to $Y$ we obtain   $\langle|b^2|\rangle\,\propto\,Y$.
  In the parton model the typical $p_T$ is independent of $Y$ and, therefore, we see that diffusion in $b$ leads to $\alpha_\pom \,\neq \,0$. However, in QCD 
  average $p_T$ depends on $Y$. Such dependence stems from the diffusion in $ \ln  p_T$ which means that in each emission of gluons $\ln p_T$ changes by a constant. Being a general features of all theories with dimensionless coupling such  diffusion in $\ln p_T$  comes out from the BFKL equation leading to $\langle \ln^2 \Lb p^2_T/p^2_{0.T}\Rb \,=\,4 D_0 \,Y$ (see \eq{DIFF}).  From this formula one can see that we see two different branches in the BFKL equation: one leads to a rapid increase of the typical transverse momentum while another to a steep decrease. This decrease does not influence the calculation of the average $p_T$, resulting in $\alpha'_\pom = 0$ for the BFKL equation. Modeling confinement by introducing cutoff in the sizes of produced dipoles we prohibit the decrease of the typical transverse momenta of the emitted gluon. As a result, the only diffusion in the large transverse momenta occurs leading to negligible $\alpha'_\pom$.

\begin{figure}
    \begin{minipage}{10.7cm}{
  \leavevmode
      \includegraphics[width=10cm]{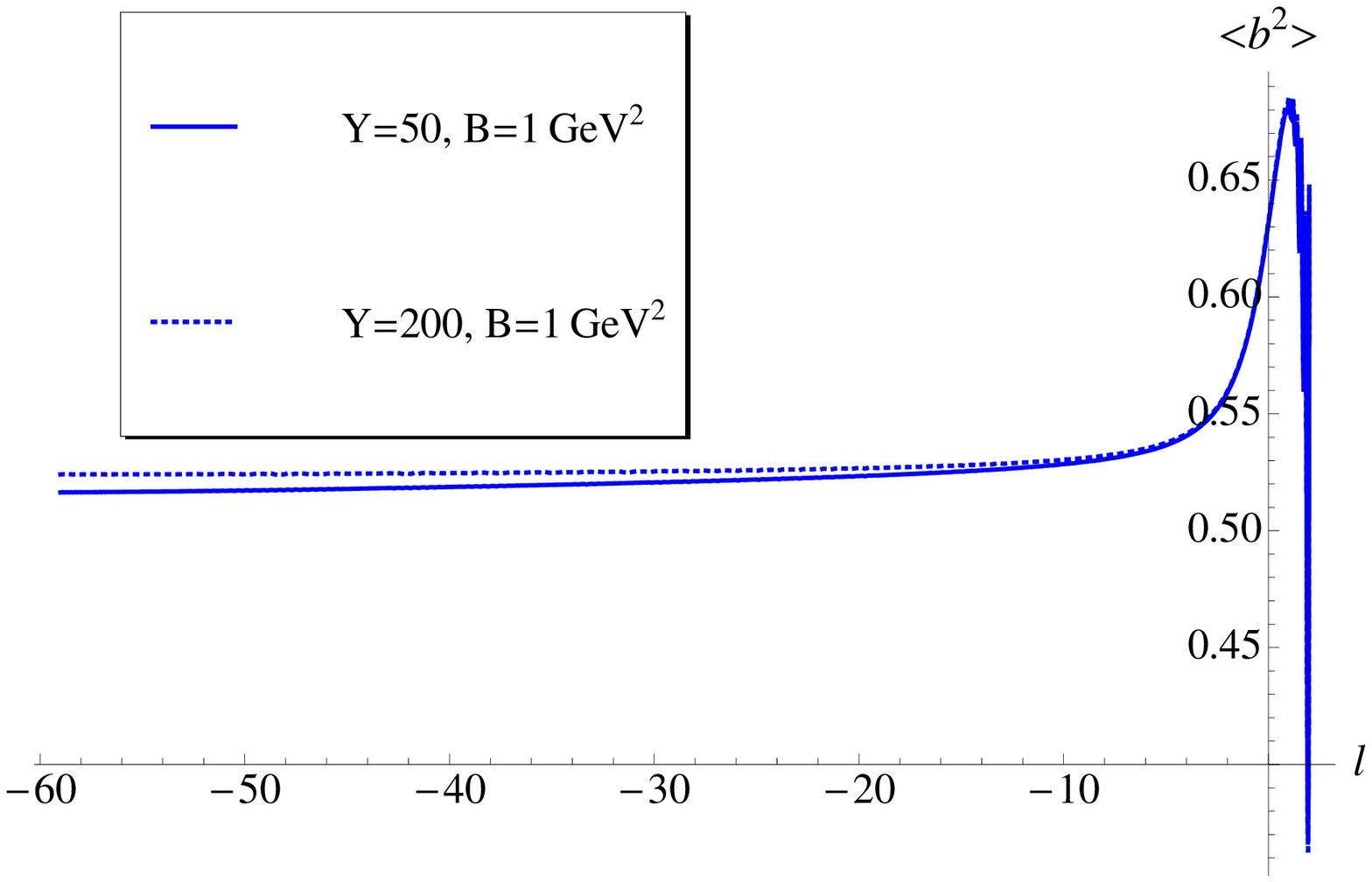}  }
      \end{minipage}
      \begin{minipage}{6.7cm}{\caption{ $\langle|b^2\Lb Y,l\Rb|\rangle$ versus $l = \ln\Lb x^2_{12}\,B\Rb$  at different values of $Y$. Solid line describes $Y=50$ while the dotted one corresponds to $Y=200$. $\protect\bas = 0.2$.}
\label{BvsL}}
\end{minipage}
   \end{figure}

\fig{BvsL} shows the dependence of $\langle|b^2\Lb Y,l\Rb|\rangle$ on $l =\ln\Lb x^2_{12}\,B\Rb$ at $B=1\,GeV^2$. We can see that the dependence of $\langle|b^2\Lb Y,l\Rb|\rangle$ on $l$ is rather weak except the region of $l$ close to 0. The maximum at $l=0$ stems from Gribov's diffusion  during the first several emissions until the average $p_T$ grows to a considerable value.


\subsection{Saturation momentum}
We have demonstrated that the modified Pomeron has a correct behaviour at large $b$ but violates the $s$-channel unitarity (Froissart theorem\cite{FROI})
both for the  partial amplitudes and for the total cross section, since they are proportional to $s^{\om_{\mbox{\tiny {BFKL}}}}$. Therefore, we need to develop the CGC/saturation approach\cite{GLR,MUQI,MV,REV} and reference therein),  based on the modified BFKL Pomeron to obtain the amplitude that will  satisfy  the unitarity constraints. We are going to develop such an approach but in this paper we wish to use the well known feature of the CGC/saturation approach: the energy behaviour of the new dimensional scale (saturation moment) can be found from the linear equation (see Refs.\cite{GLR,MUTR,MUPE}). This scale is the solution of the equation\footnote{In the first preprint version of this paper the equation for the saturation scale was written incorrectly as ${\cal N}^{BFKL}\Lb \frac{2}{Q_s\Lb Y\Rb}; Y \Rb\,\,=\,\,{\cal N}_0\,\,\leq\,\,1$. Our result for the saturation scale given in this version should be disregarded.}
\beq \label{QSEQ}
N^{BFKL}\Lb \frac{2}{Q_s\Lb Y\Rb}; Y \Rb\,\,=\,\, \frac{4}{Q^2_s\Lb Y\Rb}\,{\cal N}^{BFKL}\Lb \frac{2}{Q_s\Lb Y\Rb}; Y \Rb\,\,=\,\,{\cal N}_0\,\,\leq\,\,1
\,\,\,\,~~\mbox{where}~~~~~~~~~{\cal N}_0\,\,=\,\,\mbox{Const}
\eeq

\begin{figure}
    \begin{tabular}{c c}
  \leavevmode
      \includegraphics[width=9cm]{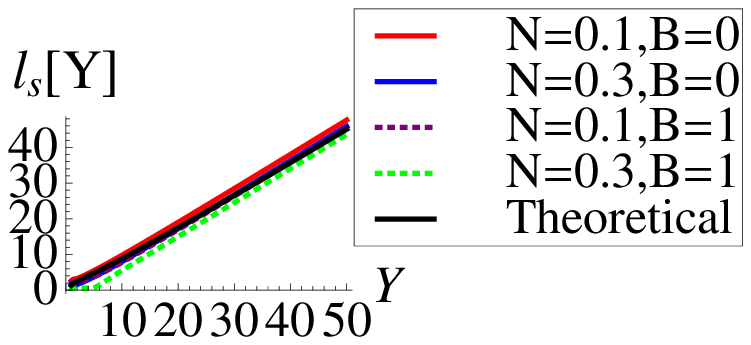}  &     \includegraphics[width=9cm]{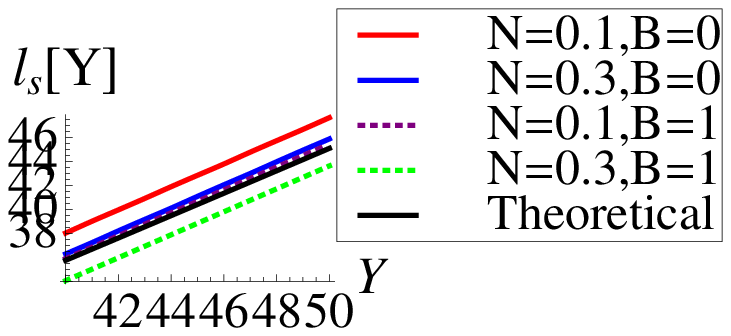} \\
      \fig{qs}-a &\fig{qs}-b\\
            \end{tabular}
\caption{ $l_s\Lb Y\Rb \,=\,\ln\Big(Q^2_s(Y)\Big{/ }Q^2_s(Y_0)\Big) $ , where $Q_s\Lb Y\Rb$ is the solution to \protect\eq{QSEQ}, versus $Y$ for different values of ${\cal N}_0$.~$\bas = 0.2$. The red solid and dotted lines correspond to solution of the BFKL equation, the solid dark brown line describes \protect\eq{QS} while the blue line is the solution to \protect\eq{QSEQ} for the modified BFKL equation. 
\label{qs}}
   \end{figure}

For the BFKL equation the solution to \eq{QSEQ} is known. It takes the form
\bea \label{QS}
&&l_s\Lb Y\Rb\,\,\equiv\,\,\ln\Big(Q^2_s(Y)\Big{/ }Q^2_s(Y_0)\Big)\,\,=\\
&&\frac{\omega(\gamma_{cr})}{1 - \gamma_{cr}}\,\,(Y - Y_0)\,\,
- \,\,\frac{3}{2 ( 1 -
\gamma_{cr})}\,\ln(Y/Y_0)\, -  \, \frac{3}{( 1 - \gamma_{cr})^2}\,\sqrt{\frac{2\,\pi}{\omega''(\gamma_{cr})}}\,
\Big(\frac{1}{\sqrt{Y}}\,-\, 
\frac{1}{\sqrt{Y_0}}\Big)\nn
\eea
where $Y = \ln(1/x)$ is our energy variable, $\omega''(\gamma)= d^2
\omega(\gamma)/(d \gamma)^2$, the value of $\gamma_{cr}$ can be found
from the equation \cite{GLR,MUTR}:
\beq \label{GAMMACR}
\frac{\omega(\gamma_{cr})}{ 1 - \gamma_{cr}} \,\,\,=\,\,\,- \,\,\frac{d \omega(\gamma_{cr})}{ d \gamma_{cr}}\,,
\,\,\mbox{with}\,\,\,\,\omega\Lb 
\ga\Rb\,\,=\,\,\bas \chi\Lb \ga \Rb\,=\,\bas \Lb 2\,\psi\Lb 1 \Rb \,-\,\psi\Lb \ga \Rb \,-\,\psi\Lb 1 - \ga \Rb\Rb
\eeq
 where $\psi\Lb \ga \Rb \,=\,d \ln \Gamma\Lb \ga\Rb/d \ga$  and $\Gamma\Lb \ga \Rb$ is the Euler gamma function.
In \eq{QS} the first term was found in Ref.\cite{GLR}, the second in Ref.\cite{MUTR} and the third term was calculated in Ref.\cite{MUPE}.
The solutions to \eq{QSEQ} for different values of ${\cal N}_0$ are plotted in \fig{qs}. Two features are clear from \fig{qs}: \eq{QS} is in a good agreement with the numerical solutions; and the energy dependences of the saturation scales are the same for the BFKL equation and for the modified BFKL equation which includes  confinement.


\section{Conclusions}
The main goal of this paper  is to find how our assumption that the size of produced dipoles  cannot be large,  will affect the main properties of the BFKL Pomeron. 
 To achieve this goal we solved the BFKL equation with the modified kernel of \eq{MODKER}. We found out that the modified BFKL Pomeron  has the same   intercept $\Delta$ as the BFKL Pomeron ( $ \Delta_{\mbox{\tiny BFKL}}\,\,=\,\,\
\om_{\mbox{\tiny BFKL}}\,\,=\,\,4\,\ln 2 \bas$) and $\alpha'_\pom = 0$. Therefore,   the  BFKL Pomeron with the modified kernel reproduces the main features of the soft Pomeron that has been found both from N=4 SYM theory\cite{KOLI,BST1,BST2,BST3} and from the high energy Reggeon phenomenology\cite{GLM,KMR}: the large value of the Pomeron intercept ($\om_0 \approx \,0.2 - 0.3$) and  $\alpha'_\pom = 0$. These both conclusions are in agreement with the numerical  solution of the modified BFKL and BK equations \cite{BEST2,BEST1}.

We consider as one of the results of this paper that we developed several methods to solve the modified BFKL equation analytically (semi-classical and diffusion approximations). The fact that these methods work we checked with the numerical calculation.

Actually, we were surprised that the model for confinement changed so little in the BFKL Pomeron and on qualitative level, the Pomeron that emerges from the modified BFKL equation, looks quite the same at the BFKL Pomeron, both in parameters and in character of the energy behaviour.  It seems that the only difference between the BFKL Pomeron and the modified BFKL Pomeron is that the second has a correct large impact parameter behaviour. 

 We believe that this statement does not depend on the particular form of \eq{MODKER}. As it has been mentioned the analytical approaches that have been developed in sections 3.1.2-3.1.4 actually are based on the kernel in which $\exp\Lb - B x^2\Rb$ is replaced by $\Theta\Lb 1/B - x^2\Rb$. This approach led to the same properties as the kernel of \eq{MODKER}.  Considering \eq{STAT} and introducing $E = - \om$ one can see that \eq{STAT} can be viewed as 
\beq \label{C1}
E\,\Psi\Lb x_{12}, \om\Rb\,\,=\,\,{\cal H} \Psi\Lb x_{12}, \om\Rb
\eeq
where ${\cal H}$ is a Hamiltonian ( see \eq{STAT}).

At short distance the BFKL wave function $\Psi_{\tiny{BFKL}}\Lb x_{12}, \om \Rb\,\,=\,\,\Lb x^2_{12}\Rb^{- \h \pm i \nu}$ is the 
eigenfunction of \eq{C1} with the eigenvalue $\om\Lb-\h \pm i \nu,0\Rb$.  Any eigenfunction of \eq{C1} will have the following two limits:
\beq \label{C2}
\Psi\Lb x_{12}, \om\Rb\,\,\xrightarrow{x_{12} \,\to\,0}\,\,\Psi_{\tiny{BFKL}}\Lb x_{12}, \om \Rb\,\,\,\,\,\mbox{and}\,\,\,\,\,\,\,\,
\Psi\Lb x_{12}, \om\Rb\,\,\xrightarrow{x_{12} \,\to\,\infty}\,\,\mbox{Const}
\eeq
At first sight the condition at long distances will restrict the values of $\nu$ in the comparison with the BFKL equation. However, it is not the case. Indeed, the eigenvalues  of the BFKL equation is degenerate having two eigenfunctions with positive and negative $\nu$. One can see that we can  find the sum of these two eigenfunction ( $ \Psi\Lb x_{12}, \om\Rb\,\,=\,\Lb 1/x^2_{12}\Rb^\h \sin\Lb  \nu /x^2_{12}\Rb$ which tends to zero at $x_{12} \to\infty$. Therefore, at any $\nu$ we can satisfy the  second condition of \eq{C2}.
On the other hand for $\nu = i \kappa \,\,( \kappa \,>\,0)$ we have two eigenfunctions: $\Lb x^2_{12}\Rb^{- \h - \kappa}$ and $\Lb x^2_{12}\Rb^{- \h +\kappa}$ . The normalization condition selects out the only eigenfunction  $\Lb x^2_{12}\Rb^{- \h +\kappa}$  which  has no divergency at $x_{12} \to 0$. Using this function we cannot satisfy the condition at $x_{12} \to  \infty$ and, therefore, we have
no solution of \eq{C1} for $\nu = i \kappa$. Hence, we expect that the spectrum for the modified Hamiltonian will be the same as the BFKL spectrum. We plan to investigate  different models of confinement and demonstrate that our general arguments works.

The independence of the spectrum of the BFKL Pomeron on the models for the confinement  gives us a hope that the unknown confinement will
change only slightly the equations of the CGC/saturation approach and these changes will not depend on the particular way of taking into account the  long distances physics. In simple words, this paper gives a hope that the CGC/saturation approach will be still a theory in spite of needed model modifications due to confinement.  The main ingredient of the CGC/saturation approach: the saturation momentum, can be calculated from the solution of the linear equation. Its value turns out to be the same as for the BFKL equation. This fact confirms our expectations that a modification of the BFKL kernel for correct large $b$ behaviour  will not lead to a significant alteration of the CGC/saturation approach.

 \eq{EQ2} is new and using this equation we are able to calculate directly $ \langle|b^2\Lb Y,l\Rb|\rangle$ which gives the information on the effective slope of the resulting Pomeron.  However, we have not solved the modified BFKL equation at fixed $b$.  Therefore, at the moment we cannot discuss   changes that the correct $b$ behaviour could trigger in azimuthal correlations that are originated by the BFKL Pomeron. However, since we introduce a new dimensional scale $B$ that cuts large distances, we can expect  changes in the estimates of local anisotropy and density variation  outside of the saturation region ( see Kovner's talk \cite{KOVT}).  Nevertheless it is too early to discuss this topic without obtaining the solution.

Our discussions with our colleagues show that we need to comment on the BFKL Pomeron with running QCD coupling. It has been intensively discussed in Refs. \cite{KLR}  how to satisfy the general initial conditions that are originated by  confinement in the case of the BFKL Pomeron with running QCD coupling. In particular, it turns out that the confinement manifests itself in a series of the Regge poles making the entire picture close to the high energy phenomenology based on the Regge poles. However as it has been discussed in section 2, it is not enough to satisfy the initial conditions to introduce the correct impact parameter behaviour. We have to change the BFKL kernel. In the approach of Refs.\cite{KLR} the  initial conditions that stem from the confinement , are  satisfied  without making any corrections to the BFKL kernel, and this approach does not change the large impact parameter behaviour of the scattering amplitude. Thus we need to take into account the running QCD coupling in addition to the modeling of confinement in the BFKL kernel. We are planning to do this in our future publications. At the moment, it is clear that the confinement modeling provides the scale for freezing the running QCD coupling that has been introduced in all numerical solutions of the non-linear equation (see Ref.\cite{AAMS}).

We believe that this paper will be useful in the search of the theoretical motivated way to include the non-perturbative corrections at large values of the impact parameters as well  as in understanding of the main ingredients of high energy phenomenology for soft processes. In the future publication we are going to study how the way of introducing confinement into the BFKL equation could change  the features of the Pomeron and to develop the CGC/saturation approach, based on the modified BFKL Pomeron.

  
  \section{Acknowledgements}
  
 We thank our    colleagues at UTFSM and Tel Aviv university for encouraging discussions. We also thank Lev Lipatov for fruitful discussions on the subject of this paper. Our special thanks go to Marat Siddikov, who participated in all discussions and in part of calculations. His contribution was very essential for us and we  considered him as one of the authors. However, he declined this offer  on the ground  that his contribution was not sufficient. He won our respect but  now we have a problem how to express our deep gratitude to him.  This research was supported by the  Fondecyt (Chile) grant 1100648.

 \end{document}